\documentclass[a4paper,11pt]{article}
\usepackage{jcappub} 
\usepackage{lineno}

\usepackage{xparse}  

\usepackage{siunitx}  
\DeclareSIUnit\parsec{pc}

\usepackage{enumitem}
\setlist{topsep=1ex,itemsep=0ex,partopsep=1ex,parsep=1ex}

\newcommand{\mH}{m_\text{H}}
\newcommand{\gH}{g_\text{H}}
\newcommand{\nH}{n_\text{H}}
\newcommand{\rhoH}{\rho_\text{H}}

\newcommand{\pH}{p_\text{H}}
\newcommand{\hatnH}{\hat{n}_\text{H}}

\newcommand{\NH}{N_\text{H}}

\newcommand{\OmegaH}{\Omega_\text{H}}

\newcommand{\tform}{t_i}

\newcommand{\aRH}{a_\text{RH}}
\newcommand{\TRH}{T_\text{RH}}
\newcommand{\mbhi}{
M_{{\rm BH},i}
}

\newcommand{\rhoBH}{\rho_\text{BH}}
\newcommand{\rhoR}{\rho_\text{R}}
\newcommand{\rhotot}{\rho_\text{tot}}
\newcommand{\nBH}{n_\text{BH}}
\newcommand{\hatnBH}{\hat{n}_\text{BH}}

\newcommand{\be}{\begin{eqnarray}}
\newcommand{\ee}{\end{eqnarray}}

\newcommand{\mpl}{ M_{\rm Pl}  }
\newcommand{\nbh}{  n_{\rm BH}  }

\newcommand{\tbh}{ T_{\rm BH} }
\newcommand{\taubh}{ \tau_{\rm BH} }

\newcommand{\mbh}{  M_{\rm BH}  }

 \newcommand{\brac}[2]{ \left( \frac{#1}{#2} \right) } 

\newcommand{\pad}[1]{{#1}\mkern2mu\vphantom{#1}}
\newcommand{\tildet}{\pad{\tilde{t}}}

\newcommand{\totalenergy}{\mathcal{E}}

\newcommand{\Neff}{N_\text{eff}}

\renewcommand{\d}{d}
\NewDocumentCommand{\dd}{o m g g}{\frac{\d\IfValueT{#1}{^#1}\IfValueT{#3}{#2}}{\d\IfValueTF{#3}{#3}{#2}\IfValueT{#1}{^#1}\IfValueT{#4}{\d #4}}}
\NewDocumentCommand{\ddt}{o g}{\IfValueTF{#2}{\dd[#1]{#2}{t}}{\dd[#1]{t}}}

\newcommand{\defeq}{\equiv}
\newcommand{\mean}[1]{\left\langle #1 \right\rangle}

\title{\boldmath Warm Hawking Relics From 
\\ Primordial 
Black Hole Domination}

\author[a]{Christopher J.~Shallue}
\affiliation[a]{Center for Astrophysics \textbar\ Harvard \& Smithsonian, Cambridge, MA 02138}
\author[b]{Julian B.~Mu\~noz}
\affiliation[b]{Department of Astronomy, The University of Texas at Austin, Austin, TX 78712}
\author[c,d,e]{Gordan Z. Krnjaic}
\affiliation[c]{Theoretical Physics Division, Fermilab, Batavia, Illinois 60510}

\affiliation[d]{Department of Astronomy and Astrophysics, University of Chicago, Chicago, IL 60637}
\affiliation[e]{Kavli Institute for Cosmological Physics, University of Chicago, Chicago, IL 60637}

\emailAdd{cshallue@cfa.harvard.edu}

\abstract{
We study the cosmological impact of warm, dark-sector relic particles produced as Hawking radiation in a primordial-black-hole-dominated universe before big bang nucleosynthesis.
If these dark-sector particles are stable, they would survive to the present day as \textit{Hawking relics} and modify the growth of cosmological structure. 
We show that such relics are produced with  much larger momenta, but in smaller quantities than the familiar thermal relics considered in standard cosmology.
Consequently, Hawking relics with keV--MeV masses affect the growth of large-scale structure in a similar way to eV-scale thermal relics like massive neutrinos.
We model their production and evolution, and show that their momentum distributions are broader than comparable relics with thermal distributions.
Warm Hawking relics affect the growth of cosmological perturbations and we constrain their abundance to be less than $2\%$ of the dark matter over a broad range of their viable parameter space.
Finally, we examine how future measurements of the matter power spectrum can distinguish Hawking relics from thermal particles.
}
\begin{document}
\maketitle
\flushbottom

\section{Introduction}\label{sec:intro}

The earliest moments of the universe remain beyond the reach of direct observations. The first era with observable consequences is big bang nucleosynthesis (BBN), when the first atomic nuclei formed a few minutes into cosmic history. Before nucleosynthesis, there ought to be a period of \textit{reheating}, in which the energy that dominated the universe during inflation was transformed into a hot plasma containing the Standard Model particles that eventually formed the first nuclei. When and how reheating took place are key open questions of modern particle cosmology \cite{allahverdi_ReheatingInflationaryCosmology_2010,amin_NonperturbativeDynamicsReheating_2015,allahverdi_FirstThreeSeconds_2021}.
An intriguing possibility is that the early universe underwent a phase in which its energy density was dominated by primordial black holes, which then evaporated through Hawking radiation~\cite{hawking_ParticleCreationBlack_1975}.
This is a generic consequence of ultra-light primordial black holes (with masses $M \lesssim \SI{e8}{\gram}$), as even a small initial abundance of such objects would quickly come to dominate the universe as it expanded \cite{barrow_BaryogenesisExtendedInflation_1991a}.
Given their short lifetimes ($< \SI{0.1}{\s}$), these black holes would evaporate to reheat the universe before nucleosynthesis, making it difficult to distinguish this scenario from other models of reheating. 

Since Hawking radiation produces all kinematically available particle species, evaporating primordial black holes may have populated the universe with particles beyond the Standard Model. If any of these particles are stable and persist to the present day, we call them \textit{Hawking relics}.
Massless Hawking relics would contribute to the cosmic radiation budget---conventionally parameterized by the number of effective neutrino species, $\Neff$---and could be detected in measurements of the cosmic microwave background~\cite{hooper_DarkRadiationSuperheavy_2019,hooper_HotGravitonsGravitational_2020,lunardini_DiracMajoranaNeutrino_2020,masina_DarkMatterDark_2020,Domenech:2021wkk,arbey_PrecisionCalculationDark_2021,cheek_EvaporationPrimordialBlack_2022,cheek_RedshiftEffectsParticle_2022}.
However, massless relics are a generic feature of many beyond-Standard-Model theories, so a deviation from the Standard Model prediction of $\Neff=3.045$~\cite{mangano_RelicNeutrinoDecoupling_2005a,desalas_RelicNeutrinoDecoupling_2016a} would not strongly favor Hawking relics over scenarios.
By contrast, massive Hawking relics can become non-relativistic in the late universe and consequently affect large-scale structure formation as part (or all) of the dark matter. In this paper, we demonstrate that massive Hawking relics leave observable signatures that can, in principle, distinguished from other beyond-Standard-Model theories.

{Massive Hawking relics can possess non-negligible thermal velocities and therefore affect structure formation in a similar way to other types of warm or hot matter, such as massive neutrinos~\cite{desalas_RelicNeutrinoDecoupling_2016a,abazajian_SterileNeutrinoHot_2001,dolgov_MassiveSterileNeutrinos_2002,giudice_StandardModelNeutrinos_2001,primack_HotDarkMatter_2001}, neutralinos~\cite{hisano_NeutralinoWarmDark_2001}, and other warm dark matter candidates~\cite{bode_HaloFormationWarm_2001,viel_WarmDarkMatter_2013,lovell_PropertiesWarmDark_2014,munoz_EfficientComputationGalaxy_2018,deporzio_FindingEVscaleLight_2021,xu_CosmologicalConstraintsLight_2022,banerjee_SignaturesLightMassive_2022}. Unlike thermally-produced relics, which have comparable number densities to the relic photons of the cosmic microwave background (CMB), Hawking relics are far less numerous.
As a consequence, Hawking relics are allowed to be orders of magnitude heavier than the upper bound of $\sim \SI{10}{\eV}$ on thermal relics \cite{xu_CosmologicalConstraintsLight_2022}.}
Moreover, Hawking relics have non-thermal ``graybody" momentum distributions, giving them a distinct imprint on the large-scale structure of the universe.
We compute the precise momentum distribution of Hawking relics, accounting for the difference in the instantaneous emission spectrum from a thermal distribution, the changing black-hole temperature during evaporation, and the redshifting of previously emitted particles as the universe expands during evaporation. We then study how Hawking relics affect different cosmic observables, including the cosmic microwave background and the large-scale structure of the universe, and use these observables to derive new constraints on warm Hawking relics.

Previous studies have used similar techniques to constrain the particle mass, as well as the masses and initial abundance of primordial black holes, under the assumption that a Hawking relic accounts for \textit{all} of the dark matter
\cite{fujita_BaryonAsymmetryDark_2014,allahverdi_NonthermalProductionDark_2018,lennon_BlackHoleGenesis_2018,morrison_MelanopogenesisDarkMatter_2019,baldes_NoncoldDarkMatter_2020,gondolo_EffectsPrimordialBlack_2020,masina_DarkMatterDark_2020,auffinger_BoundsWarmDark_2021,masina_DarkMatterDark_2021,cheek_PrimordialBlackHole-I-SolelyHawking_2022,cheek_EvaporationPrimordialBlack_2022,cheek_RedshiftEffectsParticle_2022,Haque:2024cdh}.
In this scenario, the particle mass must exceed $\sim \SI{10}{\MeV}$ to avoid suppressing structure formation beyond observational limits \cite{cheek_RedshiftEffectsParticle_2022}. For a particle this heavy to match (rather than exceed) the observed dark matter density, either
\begin{enumerate}
    \item the initial abundance of primordial black holes must be sufficiently tiny (e.g., if the particle mass is \SI{1}{\GeV}, the fraction of cosmic energy density that formed black holes cannot exceed $\num{e-9}$ \cite{cheek_RedshiftEffectsParticle_2022}); or
    \item the particle must be so massive ($\gtrsim \SI{e9}{\GeV}$ \cite{hooper_DarkRadiationSuperheavy_2019,cheek_PrimordialBlackHole-I-SolelyHawking_2022}) that its production is suppressed until the end of black-hole evaporation.
\end{enumerate}
Several other studies have considered dark matter produced by primordial black hole evaporation plus an additional mechanism, such as thermal freeze-in/freeze-out \cite{gondolo_EffectsPrimordialBlack_2020,bernal_DarkMatterTime_2021,bernal_GravitationalDarkMatter_2021,bernal_SuperradiantProductionHeavy_2022,cheek_PrimordialBlackHole-II-Interplay_2022a} or stable black hole remnants \cite{Haque:2024eyh,Barman:2024iht}.
The Hawking relics we consider in this paper do not necessarily account for all the dark matter. These relics are allowed to be warm or hot because the rest of the dark matter would still be cold: the relic would only produce a fractional suppression of the matter power spectrum.
Refs.~\cite{Chen:2023lnj,Chen:2023tzd} recently studied sterile neutrinos produced as Hawking relics and calculated their observational signatures in X-rays and gravitational waves. In this study, we investigate generic Hawking relics with different spins, focusing primarily on their signatures in the matter power spectrum. We mainly consider the case where primordial black holes dominate the energy density of the early universe, a scenario known as \textit{black-hole domination} \cite{barrow_BaryogenesisExtendedInflation_1991a}. As we will discuss in Section~\ref{sec:background-evolution}, this is a fairly generic outcome even if the initial abundance of primordial black holes is small, and it leads to a universal attractor solution of the spacetime dynamics, making it a natural case to study.

Our key finding is that warm Hawking relics can induce changes in the matter power spectrum that are observable by current spectroscopic galaxy surveys like SDSS~\cite{SDSS:2006lmn}, DESI~\cite{desicollaboration_DESIExperimentPart_2016a}, HETDEX~\cite{gebhardt_HobbyEberlyTelescopeDark_2021}, and PFS~\cite{takada_ExtragalacticScienceCosmology_2014}. Moreover, these Hawking relics can be distinguished, in principle, from other kinds of relics (such as those produced by thermal freeze-out) due to their unique graybody momentum distributions.
We derive new constraints on warm Hawking relics, finding that the largest allowed particle mass is $\sim \SI{50}{\MeV}$ in the black-hole-domination scenario, about six orders of magnitude larger than the heaviest possible thermally-produced relic \cite{xu_CosmologicalConstraintsLight_2022}. We constrain the abundance of warm Hawking relics to be less than $\sim 2\%$ of dark matter, even if primordial black holes produced multiple different kinds of relic particles. Finally, we find that for a large region of parameter space, Hawking relics behave like thermal relics with temperatures that are independent of the mass of particle and the initial mass and abundance of the black holes. These temperatures depend only on the particle's spin and number of degrees of freedom, so if a future investigation finds evidence for a thermal relic at one of these temperatures, further investigation of its momentum distribution will be warranted to determine whether it could be a Hawking relic.

This paper is structured as follows. In Section~\ref{sec:preliminaries}, we review the key background concepts needed for the rest of the paper. In Section~\ref{sec:background-evolution}, we study the dynamics of spacetime with primordial black holes and work out the details of reheating. In Section~\ref{sec:mom-distribution}, we calculate the momentum distribution of a Hawking relic and investigate its dependence on the changing black-hole temperature and the expansion of the universe. In Section~\ref{sec:observing}, we consider the observable consequences of warm Hawking relics in the black-hole-domination scenario. We evolve the momentum distribution of a Hawking relic to the present day and use it to compute measurable effects on the cosmic microwave background and large-scale structure of the universe. We use \texttt{CLASS} \cite{blas_CosmicLinearAnisotropy_2011} to compute the matter power spectrum in the presence of Hawking relics of different masses, and study the differences and similarities between Hawking relics and thermal, neutrino-like relics. Finally, we derive new constraints on warm Hawking relics by approximating a Hawking relic by an ``equivalent'' thermal fermion and using constraints on thermal fermions from a joint analysis of CMB, galaxy clustering, and weak-lensing data from Ref.~\cite{xu_CosmologicalConstraintsLight_2022}. In Section~\ref{sec:conclusion}, we summarize our main results and give concluding remarks.

Throughout this paper we use units with $c=\hbar=k_B=1$. Our \texttt{CLASS} configuration uses the following $\Lambda$CDM cosmological parameters: $\Omega_b h^2 =  0.0223$, $\Omega_c h^2 = 0.120$, $h = 0.678$, $A_s = \num{2.10e-09}$, $n_s = 0.966$, $\tau_\text{reio.} = 0.0543$, in agreement with Ref.~\cite{aghanim_Planck2018Results_VI_2020}.

\section{Primordial Black Hole Preliminaries}\label{sec:preliminaries}

\subsection{Formation}\label{sec:PBH-formation}

Various cosmological scenarios predict the formation of black holes in the early universe. 
For example, primordial black holes may have formed from the gravitational collapse of overdense regions \cite{zeldovich_HypothesisCoresRetarded_1967,carr_BlackHolesEarly_1974}.
In a radiation-dominated era, the critical overdensity needed for a region to collapse into a black hole is $\delta_c \gtrsim 0.4$ \cite{carr_PrimordialBlackHole_1975,carr_ConstraintsPrimordialBlack_2021}.
Densities this high would be extremely rare according to the standard theory of nearly scale-independent small density perturbations generated during inflation. However, the primordial power spectrum has only been measured out to wavenumbers of order $k \sim \SI{10}{\per\mega\parsec}$ \cite{chabanier_MatterPowerSpectrum_2019,sabti_NewRoadsSmallscale_2022}, leaving open the possibility of small-scale, high-amplitude density perturbations in the early universe that produced a meaningful population of black holes.
Such perturbations arise, for example, in some models of single- and multi-field inflation \cite[][and references therein]{geller_PrimordialBlackHoles_2022,qin_PlanckConstraintsGravitational_2023,ozsoy_InflationPrimordialBlack_2023}.
Alternatively, primordial black holes may have formed due to a sudden reduction in pressure during a phase transition, or through more exotic scenarios.
See Ref.~\cite{carr_ConstraintsPrimordialBlack_2021} for a recent review of primordial black hole formation scenarios and constraints.

If primordial black holes formed from the gravitational collapse of primordial fluctuations, their masses are expected to be of order $\mbhi \sim M_H$, where $\mbhi$ is the formation mass of a given black hole (so denoted to differentiate from its time-dependent mass $\mbh(t)$ due to evaporation) and $M_H$ is the mass in a Hubble volume at its formation time.
This is because overdense regions are stretched to super-horizon scales during inflation and only undergo collapse when they re-enter the horizon. It is conventional to parameterize $\mbhi = \gamma M_H$, where $\gamma$ is an order-unity parameter that depends on the details of gravitational collapse \cite{carr_PrimordialBlackHole_1975,carr_NewCosmologicalConstraints_2010}. We then have
\begin{equation}\label{eq:Mform-rhoform-tform}
    \mbhi
    = \frac{\gamma}{2GH} \sim  10^4 \, 
 {\rm g}  \brac{ \gamma}{0.1} \brac{10^9 \, \rm GeV}{H},
\end{equation}
where $H$ is the Hubble rate at formation.

We will assume throughout this paper that all primordial black holes formed at approximately the same time and with approximately the same mass. This so-called monochromatic mass spectrum is commonly assumed in the primordial black hole literature, although the true spectrum depends on the details of primordial black hole formation and could be extended even if primordial black holes arose from a narrow peak in the power spectrum \cite{carr_PrimordialBlackHole_1975,yokoyama_CosmologicalConstraintsPrimordial_1998a,carr_ConstraintsPrimordialBlack_2016,carr_PrimordialBlackHoles_2016}. See Ref.~\cite{cheek_EvaporationPrimordialBlack_2022} for a recent investigation of dark matter produced by evaporating primordial black holes with extended mass distributions.
We also neglect black-hole spin, as our focus is on ultra-light black holes ($\mbhi \lesssim \SI{5e8}{\gram}$), which are not expected to form with significant spin or to acquire it through accretion~\cite{luca_InitialSpinProbability_2019,luca_EvolutionPrimordialBlack_2020}. If primordial black holes possessed significant spin, their production of higher-spin particles would be enhanced~\cite{page_ParticleEmissionRates-II-MasslessRotating_1976}, which would modify the predicted distribution of relic particles~\cite{hooper_HotGravitonsGravitational_2020,arbey_PrecisionCalculationDark_2021,masina_DarkMatterDark_2021,cheek_PrimordialBlackHole-I-SolelyHawking_2022,cheek_PrimordialBlackHole-II-Interplay_2022a,cheek_RedshiftEffectsParticle_2022,cheek_EvaporationPrimordialBlack_2022}.

\subsection{Constraints}

The Hubble rate after inflation is bound by Planck constraints on the tensor-to-scalar ratio to satisfy $H \lesssim 10^{14}$ GeV \cite{akrami_Planck2018Results_X_2020}.
By Eq.~\eqref{eq:Mform-rhoform-tform}, the corresponding bound on black holes that formed from gravitational collapse is approximately $\mbhi \gtrsim 1 g$, where we have taken $\gamma \sim 1$, so we do not consider formation masses below this threshold.
We are only interested in primordial black holes that evaporate by the present day, which true if $\mbhi \lesssim \SI{e15}{\gram}$. However, if there were a meaningful number of evaporating black holes during big bang nucleosynthesis (BBN), the predicted abundances of atomic nuclei would be incorrect \cite{carr_ConstraintsPrimordialBlack_2021}. We thus require that the primordial black holes evaporate before BBN, which gives us an upper bound of $\mbhi \lesssim \SI{5e8}{\gram}$ \cite{Keith:2020jww} as we will show in Section~\ref{sec:reheating}. 

The abundance of evaporating primordial black holes is conventionally described by the parameter $\beta$, defined as the fraction of the cosmic energy density that collapsed into black holes,
\begin{equation}
    \beta \defeq \frac{\rhoBH(\tform)}{\rhotot(\tform)}~,
\end{equation}
where $\tform$ is the formation time ~\cite{sasaki_PrimordialBlackHoles_2018}.
For primordial black holes that evaporate before BBN, the tightest constraints on $\beta$ come from gravitational waves that would be generated during the rapid transition from matter to radiation domination as the black holes evaporate. These gravitational waves would contribute to the effective number  of relativistic species ($N_\text{eff}$) \cite{inomata_GravitationalWaveProduction_2020}, which constrains \cite{domenech_GravitationalWaveConstraints_2021} 
\begin{equation}
    \beta \lesssim 10^{-6} \left(\frac{\mbh}{\SI{e4}{\gram}}\right)^{-17/24}~,
\end{equation}
though this bound is model dependent and can be offset by additional entropy transfers into the Standard Model radiation bath after reheating from Hawking evaporation. 
There are also gravitational waves due to gravitons produced directly by Hawking radiation, whose production is significantly enhanced for black holes with high angular momentum \cite{page_ParticleEmissionRates-II-MasslessRotating_1976}.
Both kinds of gravitational waves---those generated by rapid evaporation and those directly produced by Hawking radiation---may be detectable in future measurements of the cosmic microwave background \cite{hooper_DarkRadiationSuperheavy_2019,hooper_HotGravitonsGravitational_2020,arbey_PrecisionCalculationDark_2021,Domenech:2021wkk}.

Aside from gravitational radiation, primordial black holes that evaporated before BBN are difficult to constrain because they would have evaporated into Standard Model particles and thermalized prior to our earliest observations (see Appendix~\ref{appendix:thermalization}). This paper explores a way to detect these hypothetical black holes using Hawking relic particles that do not thermalize and stream freely to the present day.

\subsection{Evaporation}\label{sec:hawking-radiation}

Observers far from a black hole will measure a flux of particles from the direction of the black hole \cite{hawking_ParticleCreationBlack_1975}. This phenomenon, known as Hawking radiation, is a theoretical consequence of adding quantum fields to a curved black-hole spacetime.
Although originally derived for astrophysical black holes formed from stellar collapse, the same effect should also apply to primordial black holes formed in the early universe. We will assume that Hawking radiation from primordial black holes operates in the same way as black holes in a vacuum. Through Hawking radiation, black holes are expected to lose mass and eventually evaporate.\footnote{A complete dynamical model of Hawking evaporation requires a theory of quantum gravity, so the full details of this process are currently uncertain, especially in the later stages of the black-hole lifetime \cite{dvali_BlackHoleMetamorphosis_2020}. For example, evaporating black holes might leave behind Planck-scale remnants, which could themselves be dark-matter candidates \cite{macgibbon_CanPlanckmassRelics_1987,barrow_CosmologyBlackHole_1992,carr_BlackHoleRelics_1994}. However, in this paper we assume that Hawking evaporation proceeds until black holes evaporate completely.}  

For a Schwarzschild black hole of mass $\mbh$, Hawking radiation is characterized by a black-hole temperature
\be
\tbh = \frac{1}{8\pi G\mbh}
\approx 10^{13} \, {\rm GeV}\brac{\rm g}{\mbh}.
\ee
The instantaneous rate of particles emitted by Hawking radiation as measured by distant observers at late times is
\begin{equation}\label{eq:dNdEdt-Gamma}
    \dd{N_{j}}{E}{t} = \frac{g_j}{2\pi} \frac{\Gamma_{j}}{e^{E/T_\text{BH}} \pm 1}~~, ~
\end{equation}
where $N_j$ is the number of  particles of species $j$ emitted, $E$ is their energy and $g_j$ their number of degrees of freedom, and the $\pm$ characterizes fermions or bosons \cite{page_ParticleEmissionRates-I-MasslessNonrotating_1976}. Here $\Gamma_{j}(E, \mbh)$ is the so-called graybody factor \cite{page_ParticleEmissionRates-II-MasslessRotating_1976}, which accounts for dynamical interactions between the black hole and the field of emitted particles (see Appendix~\ref{appendix:graybody}).\footnote{Throughout this paper, we use the graybody factors from Ref.~\cite{arbey_BlackHawkV2Public_2019}.}

A particle species of mass $m$ will be emitted relativistically if  $m \ll \tbh$, and for
\be\label{eq:m-upperbound}
m \ll 
\frac{1}{8 \pi G M_{{\rm BH},i}}  \approx
\SI{e13}{\GeV}  
\brac{\rm g}{\mbhi} ,
\ee
this is true for the entire lifetime of the black hole.\footnote{If there are ultra-massive dark-sector particles in nature, their production will be suppressed until the end of the black-hole lifetime and their high mass will make them cold dark matter at the present day. 
In this paper, we consider only particles emitted relativistically throughout the black-hole lifetime, in which case any dark-sector relics will be warm at the present-day.} In this regime, $\Gamma_j$ is a function of $p / \tbh$, where $p$ is the particle momentum, and $\Gamma_j$ is the same for all particles with the same spin \cite{page_ParticleEmissionRates-I-MasslessNonrotating_1976}. Then the power emitted in each species is
\begin{equation}
    \ddt{\totalenergy_j}
    = \frac{g_j}{2\pi} \int_0^\infty
\frac{ dp \, p \, \Gamma_{j}}{e^{p/T_\text{BH}} \pm 1}~~
    \equiv \frac{g_j}{(G\mbh)^2} f_j~,
\end{equation}
where we have defined the emission per degree of freedom as
\begin{equation}\label{eq:fj}
    f_j \equiv
    \frac{(G \mbh)^2}{2\pi
    } \int_0^\infty  
\frac{dp \, p\, \Gamma_{j}}{e^{p/T_\text{BH}} \pm 1}
    \approx
    \num{7.44e-5} \times
    \begin{cases}
        \num{1.00}, & \text{spin $0$}, \\
        \num{0.550}, & \text{spin $1/2$}, \\
        \num{0.226}, & \text{spin $1$},  \\
        \num{0.0258}, & \text{spin $2$.}
    \end{cases}
\end{equation}
Therefore, the black hole loses mass at a rate of 
\begin{equation}\label{eq:bh-mass-loss-rate}
    \ddt{\mbh}
    = - \sum_j \frac{{d\cal E}_j}{dt}
    = - \frac{1}{(G\mbh)^2} \sum_j g_j f_j 
    = -\frac{\kappa}{\mbh^2},
\end{equation}
where we have defined the quantity
\begin{equation}\label{eq:alpha}
    \kappa \defeq G^{-2} \sum_j g_j f_j~~,
\end{equation}
with the sum over all particle species $j$ produced by Hawking radiation. Solving Eq.~\eqref{eq:bh-mass-loss-rate} yields the black-hole mass as a function of time
\begin{align}\label{eq:M(t)}
    \mbh(t)
    = \mbhi \left(1 - \frac{t-\tform}{\taubh} \right)^{1/3},
\end{align}
where $\tform$ is the formation time, $M_{{\rm BH},i} \equiv \mbh(t_i)$ is the formation mass, and
\begin{equation}\label{eq:tau}
    \taubh \defeq \frac{M_{{\rm BH},i}^3}{3 \kappa} \sim 1 {\rm s}\, \brac{\kappa_{\rm SM}}{\kappa} \brac{{\mbh}_i}{10^9 \,\rm g}^3 
\end{equation}
is the primordial black hole lifetime, where $\kappa_{\rm SM} \approx 8 \times 10^{26} {\rm g}^3 {\rm s^{-1}}$ is the value of $\kappa$ assuming Hawking evaporation due to all Standard Model particles and no dark-sector ones.
Although the precise value of $\kappa$ depends on the full spectrum of particles in nature, which may include beyond-Standard-Model (BSM) states, any BSM contributions would only increase $\kappa$ by a few percent for each new species. Throughout this paper we make the approximation that $\kappa \approx \kappa_{\rm SM}$, which does not significantly affect our results unless there are many BSM degrees of freedom.

Since the primordial black holes we study formed during early radiation domination, we can combine Eqs.~\eqref{eq:Mform-rhoform-tform} and~\eqref{eq:tau} to obtain
\begin{equation}
    \frac{\tform}{\taubh}
    \sim \num{e-11} \left(\frac{\kappa}{\kappa_\text{SM}}\right) \left(\frac{\mbhi}{\si{\gram}}\right)^{-2} \ll 1,
\end{equation}
so to a very close approximation Eq.~\eqref{eq:M(t)} simplifies to
\begin{equation}\label{eq:M(t)-approx}
    \mbh(t) = M_{{\rm BH},i} \left(1 - \frac{t}{\taubh} \right)^{1/3},
\end{equation}
which we will use throughout the rest of the paper.

\section{Spacetime Dynamics with Primordial Black Holes}\label{sec:background-evolution}

Assuming that primordial black holes form and evaporate as outlined in Section~\ref{sec:preliminaries}, we now move to study their effect on cosmic expansion and the subsequent reheating of the universe after their evaporation.

\subsection{Evolution of Energy Densities}\label{sec:friedmann-fluid}

Adding primordial black holes to the standard cosmological model requires us to modify the dynamics of the early universe.
On large scales, the universe after black-hole formation can be described as a Friedmann-Lemaître-Robertson-Walker universe with energy density consisting of (i) black holes, which behave like pressureless matter, and (ii) relativistic species, including products from Hawking radiation.
Then, the Friedmann equation is
\be
\label{eq:friedmann}
    H^2 = \frac{8 \pi G}{3} \left( \rhoBH + \rhoR  \right),
\ee
where $H$ is the Hubble parameter,  $\rhoBH$ is the energy density in black holes, and $\rhoR$ is the energy density in relativistic species. The energy densities are governed by
\be\label{eq:fluid}
    \dot{\rho}_\text{BH} = -3 H \rhoBH + \rhoBH \frac{\dot{M}_{\rm BH}}{\mbh},~~~~
    \dot{\rho}_R
    = -4 H \rho_R - \rhoBH \frac{\dot{M}_{\rm BH}}{\mbh},
\ee
where in each expression the first term represents the change in energy density due to expansion and the second term represents the conversion of black-hole rest mass into relativistic energy via Hawking radiation. 
We note that some particle species contributing to $\rhoR$ may become non-relativistic during the lifetime of the black holes, but the energy density of non-relativistic species will generally be negligible compared to the relativistic species  \cite{husdal_EffectiveDegreesFreedom_2016}. The non-relativistic species may include Standard Model particles that remain coupled to photons, such as protons and electrons, and/or hypothetical dark sector particles such as weakly interacting massive particles (WIMPs) that freeze out during black-hole evaporation.

\begin{figure}[t!]
    \centering
    \includegraphics[width=\textwidth]{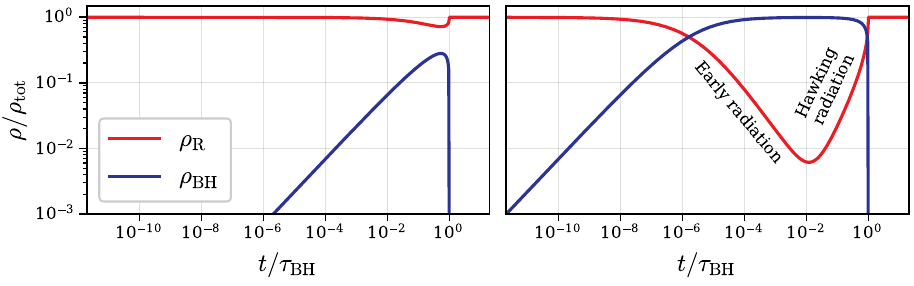}
    \caption{
    Evolution of cosmic energy density in radiation-dominated (left) and black-hole-dominated (right) scenarios.
    For any black-hole formation mass, there is a threshold value of the initial abundance $\beta$ above which black-hole domination will always occur.}\label{fig:density-evol-beta}
\end{figure}

Figure~\ref{fig:density-evol-beta} shows two possible cosmological histories if primordial black holes are present in appreciable numbers. In both cases, the black-hole formation mass is $\mbhi = \SI{e4}{\gram}$, which sets the initial cosmic energy density by Eq.~\eqref{eq:Mform-rhoform-tform}. The initial abundance of black holes is $\beta = \num{e-10}$ and $\num{e-7}$ in the left and right panels, respectively. Hawking radiation is negligible early in the black-hole lifetime, so $\rhoBH$ and $\rhoR$ initially redshift as $a^{-3}$ and $a^{-4}$, as usual for non-relativistic and relativistic matter. Eventually, Hawking radiation becomes significant and the black holes are converted into radiation. In the left panel, the initial black-hole abundance is so small that the universe remains radiation dominated throughout black-hole evaporation. In the right panel, the black-hole energy density overtakes radiation and dominates the total energy for most of the black-hole lifetime.

\begin{figure}[t!]
    \centering
    \includegraphics[width=0.75\textwidth]{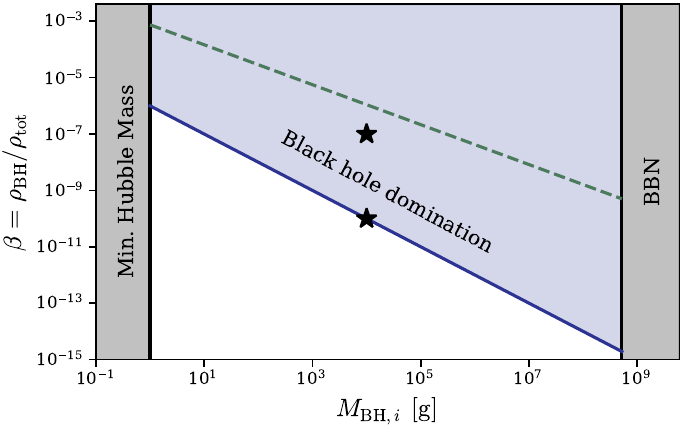}
    \caption{Primordial black hole parameter space. 
    The black-hole formation mass $\mbhi$ ranges from a lower limit of $\sim \SI{1}{\gram}$ given the minimum Hubble mass from Planck constraints on the tensor-to-scalar-ratio, to an upper limit of $\sim \SI{5e8}{\gram}$ in order for the black holes to evaporate by BBN.
    In the blue shaded region, black holes will dominate the cosmic energy density before they evaporate.
    The dashed green line is a model-dependent upper limit on the black-hole abundance from constraints on gravitational waves produced when they evaporate.
    The stars mark the parameters used in Figure~\ref{fig:density-evol-beta}.
    }\label{fig:Mform-beta-region}
\end{figure}

The scenario in the right panel of Figure~\ref{fig:density-evol-beta} is called \textit{black-hole domination}, and it is generic: for any formation mass, there is a threshold value of $\beta$ above which black-hole domination will always occur. If primordial black holes form from gravitational collapse of density fluctuations, the threshold is approximately \cite{barrow_BaryogenesisExtendedInflation_1991a}
\begin{equation}\label{eq:beta-BHD}
    \beta_\text{BHD} \approx \num{e-6} \brac{\rm g}{M_{{\rm BH},i}}.
\end{equation}
Figure~\ref{fig:Mform-beta-region} shows the region of the primordial black-hole parameter space for which black-hole domination occurs. In this region, the spacetime evolution has an attractor solution: all values of $\beta$ yield the same spacetime dynamics after black-hole domination begins. We will discuss the dynamics of black-hole domination in the following section.

\subsection{Dynamics of Black-Hole Domination}\label{sec:BHD-dynamics}

When black-hole domination occurs, the universe enters a matter-dominated phase before Hawking radiation plays a significant role, and the subsequent evolution is independent of the initial black-hole abundance $\beta$.
Since most of the Hawking radiation is emitted near the end of the black-hole lifetime, the cosmic dynamics can be approximated by a matter-dominated universe satisfying $H = 2 / (3t)$, with the black holes evaporating instantaneously at $t = \taubh$.
A more precise model of the cosmic dynamics can be obtained by numerically solving the fluid equations with appropriate initial conditions, as we describe in Appendix~\ref{appendix:background-evol-BHdom}.
Compared to the matter-domination approximation, the full numerical solution yields a smaller cosmic energy density at evaporation time, satisfying
\begin{equation}\label{eq:rhoBH_evap1}
    \rhoR(\taubh) = \frac{0.88}{6 \pi G \taubh^2} ~,
\end{equation}
where the numerator would be unity if the impact of Hawking radiation on Hubble expansion were neglected.
The comoving black-hole number density, which is constant between formation and evaporation (i.e., the number of black holes does not change), satisfies
\begin{equation}\label{eq:hat-nBH-BH-dom}
    \hatnBH
    \defeq a(t)^3 \nBH(t)
    = \frac{1.09 \aRH^3}{6 \pi G \taubh^2 \mbhi} ~,
\end{equation}
where $\aRH$ is the scale factor at reheating (evaporation time), which we will calculate in the following section. Throughout this paper, we use the full numerical solution as it is the most accurate model of the spacetime background during black-hole domination.

\subsection{Reheating Through Hawking Evaporation}\label{sec:reheating}

In the black-hole domination scenario, reheating occurs when the universe transitions from black-hole-dominated to radiation-dominated through Hawking evaporation.
We assume that all Standard Model particles emitted by black holes have reached thermal equilibrium at this time (see Appendix~\ref{appendix:thermalization} for a discussion). The total energy density at evaporation time is
\begin{equation}\label{eq:rhoevap-components}
    \rho_R = \rho_\text{SM} + \rho_\text{H} \approx \rho_{\rm SM    }~~,
\end{equation}
where $\rhoR$ is the energy density of all (initially) relativistic products of Hawking radiation, $\rho_\text{SM}$ is the energy density of species in thermal equilibrium with the Standard Model plasma, and $\rho_\text{H}$ is the energy density of non-interacting Hawking particles. If species are initially coupled to the Standard Model but decouple during primordial black hole evaporation (e.g. neutrinos or WIMPs), they are included in $\rho_\text{SM}$ not $\rho_\text{H}$. Note that due to $N_{\rm eff}$ bounds from BBN and the CMB, the non-thermalizing Hawking relics are always a subdominant fraction of the total energy density upon evaporation in any viable scenario.

The energy density is related to the reheating temperature by
\be
\label{eq:rhoBH_evap2}
\rhoR(\taubh) \approx \rho_{\rm SM}(\taubh) \equiv \frac{\pi^2}{30}  
g_{*}(T_{\rm RH})
T_{\rm RH}^4  ~,
\ee
where
$T_{\rm RH}$ is defined to be the temperature of the Standard Model at evaporation time and $g_*(T)$ is the number of relativistic degrees of freedom in equilibrium at temperature $T$.\footnote{We use the $g_*$ functions tabulated by Ref.~\cite{husdal_EffectiveDegreesFreedom_2016} with a QCD transition temperature of $T_c = \SI{214}{\MeV}$. For our purposes, the choice of $T_c$ only matters for a narrow range of black-hole formation masses for which the evaporation temperature is close to that transition.}
Combining Eqs.~\eqref{eq:rhoBH_evap1} and~
\eqref{eq:rhoBH_evap2}, we find 
\begin{equation}\label{eq:TRH}
    \TRH
    \approx \left(\frac{30 \, \rho_{R}(\taubh) }{\pi^2 g_{*,\text{RH}}}  \right)^{1/4}
      \approx \SI{2.7e10}{\GeV} \left(\frac{g_{*,\infty}}{g_{*,\text{RH}}}\right)^{1/4} \left(\frac{\si{\gram}}{M_{{\rm BH},i}}\right)^{3/2},
\end{equation}
where $g_{*,\text{RH}} \defeq g_{*}(\TRH)$, $g_{*,\infty} = 106.75$ is the value of $g_{*}(T)$ when all Standard Model particles are relativistic, and we have used Eq.~\eqref{eq:tau}.

\begin{figure}[t!]
    \centering
    \includegraphics[width=\textwidth]{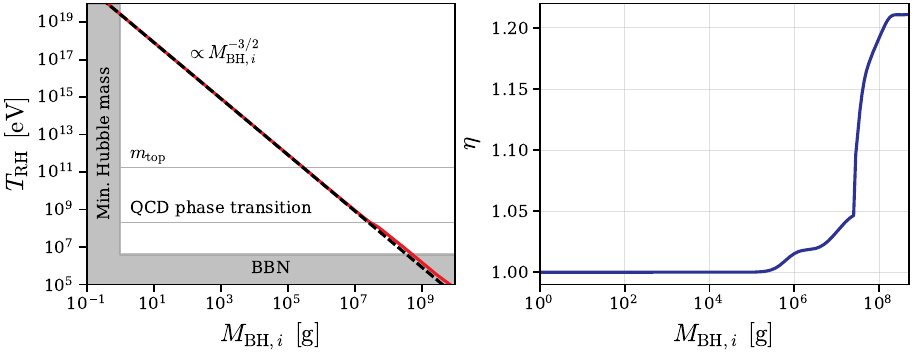}
    \caption{
    \textbf{Left:} The reheating temperature is proportional to $\mbhi^{-3/2}$ for $\mbhi \lesssim \SI{e5}{\gram}$ and deviates slightly for larger formation masses due to Standard-Model degrees of freedom becoming non-relativistic. 
    The lower bound on $\TRH$ comes from the minimum reheating temperature required for BBN, which implies an upper bound on $\mbhi$. \textbf{Right:} Correction factor $\eta$, as defined in Eq.~\eqref{eq:eta}, approaches unity for $M_{{\rm BH},i} \lesssim \SI{e5}{\gram}$ and is only slightly larger for heavier formation masses.}\label{fig:Tevap-vs-Mform}
\end{figure}

Figure~\ref{fig:Tevap-vs-Mform} shows the reheating temperature as a function of the black-hole formation mass.
Smaller black holes ($M_{{\rm BH},i} \lesssim \SI{e5}{\gram}$) evaporate while all Standard Model particles are still relativistic, which means that $g_{*,\rm RH} = g_{*,\infty}$ and $\TRH$ is exactly proportional to $M_{{\rm BH},i}^{-3/2}$.
However, larger black holes evaporate after some species have become non-relativistic, which decreases $g_{*,\rm RH}$ and heats the plasma to a higher temperature than if all particles had remained relativistic. In this case, Eq.~\eqref{eq:TRH} must be solved numerically for $\TRH$.
In order for the black holes to evaporate before BBN, at which time the plasma temperature is at most $\sim \SI{4}{\MeV}$ \cite{hannestad_WhatLowestPossible_2004,Hasegawa:2019jsa,Kawasaki:2000en}, the black-hole formation mass must satisfy $M_{{\rm BH},i} \lesssim \SI{5e8}{\gram}$ \cite{carr_NewCosmologicalConstraints_2010,carr_ConstraintsPrimordialBlack_2021,Keith:2020jww}.

Given the reheating temperature, we can also calculate the scale factor at primordial black hole evaporation. The entropy of the Standard Model plasma is conserved, so its entropy density satisfies $s \propto a^{-3}$, where $a$ is the scale factor and we adopt the convention that $a(t_0) = 1$ today. Using $s \propto g_{*,s} T^{3}$, where $g_{*,s}$ is the number of relativistic degrees of freedom in entropy, the scale factor at evaporation time satisfies
\begin{equation}\label{eq:aRH-Mform}
    a_\text{RH}
    \defeq a(\taubh)
    = \num{2.9e-24} \, \eta \left(\frac{ M_{{\rm BH},i} }{\si{\gram}}\right)^{3/2},
\end{equation}
where we set the photon temperature today to $\SI{2.725}{\kelvin}$ \cite{fixsen_CosmicMicrowaveBackground_1996}, and where we have defined
\begin{equation}\label{eq:eta}
    \eta \defeq \left(\frac{g_{*,\infty}}{g_{*,s,\text{RH}}}\right)^{1/3} \left(\frac{g_{*,\text{RH}}}{g_{*,\infty}}\right)^{1/4},
\end{equation}
where $g_{*,s,\text{RH}} \defeq g_{*,s}(\TRH)$. Figure~\ref{fig:Tevap-vs-Mform} shows $\eta$ as a function of the black-hole formation mass. It is an order unity parameter that approaches 1 for small primordial black holes ($\mbhi \lesssim \SI{e5}{\gram}$, which evaporate while all particle species are relativistic), and is slightly larger for black holes with higher formation masses. For the largest primordial black holes we consider ($\mbhi \approx \SI{5e8}{\gram}$), we find $\eta \approx 1.2$.

\section{Momentum Distribution of a Hawking Particle}\label{sec:mom-distribution}

If primordial black holes evaporated in the early universe, they would have produced all particles coupled to gravity, including any kinematically available BSM states. 
In this paper, we consider a stable, non-interacting species of particle produced by primordial black holes, which we call a \textit{Hawking particle}. We refer to remnant Hawking particles at the present day as \textit{Hawking relics}, in a similar vein to relic neutrinos. We assume that the mass $\mH$ of the Hawking particle satisfies  $m_{\rm H} \ll 10^{13} {\rm GeV} ({\rm g}/\mbhi)$ from Eq.~\eqref{eq:m-upperbound},
so that it is emitted relativistically throughout primordial black hole evaporation.
 
In this section we compute the momentum distribution of a Hawking particle. This distribution will differ from the instantaneous emission spectrum in Eq.~\eqref{eq:dNdEdt-Gamma} because (i) the black-hole temperature increases during evaporation, shifting the instantaneous emission spectrum towards higher momenta; and (ii) the universe expands during evaporation, redshifting the spectrum of previously emitted particles towards lower momenta. Both of these effects have a sizable impact on the final distribution, making it significantly wider than the instantaneous spectrum.
This affects the observability of Hawking relics, as we will explore in Section~\ref{sec:observing}.

\subsection{Static Spacetime Background}\label{sec:mom-dist-static}
\begin{figure}[t!]
    \includegraphics[width=\textwidth]{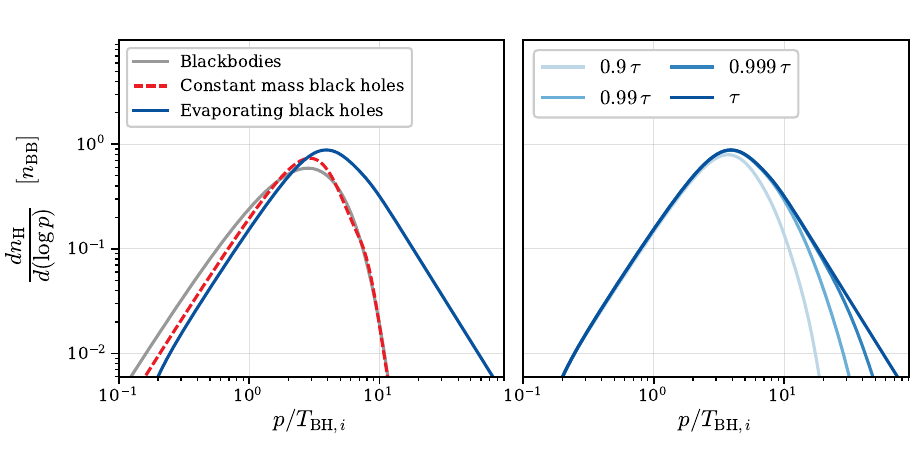}
    \caption{
    The increasing black-hole temperature during evaporation shifts the distribution of Hawking particles to higher momenta. $\nH$ is the number density of scalar particles emitted in a static universe as a function of their momenta $p$.
    \textbf{Left:} We show three scenarios for comparison. The blue line represents evaporating primordial black holes with initial mass $\mbhi$ and temperature $T_{{\rm BH}, i}$. The red line represents non-evaporating black holes, obtained from Eq.~\eqref{eq:dnHdp-static} by holding the black-hole mass constant at $\mbhi$ over the time interval $\taubh$. The gray line represents non-evaporating blackbodies with temperature $T_{{\rm BH}, i}$, obtained from Eq.~\eqref{eq:dnHdp-static} by replacing graybody factors with $\Gamma_H = 27 (G \mbhi p)^2$ (see Appendix~\ref{appendix:graybody}). The $y$-axis is in units of $n_{\rm BB}$, the total number density emitted from non-evaporating blackbodies.
    {\bf Right:} Evolution of the momentum distribution over the lifetime of the evaporating black holes. Most of the high-momentum tail is emitted in the final stages of evaporation, which is also where the theory of Hawking radiation is most uncertain \cite{dvali_BlackHoleMetamorphosis_2020}.
    }\label{fig:dndp-static}
\end{figure}

To isolate the effect of the increasing black-hole temperature on the final momentum distribution, we first consider evaporating black holes in a static spacetime background (i.e., no redshifting of particle momenta). Since Hawking particles are defined to not interact, the momentum of each particle remains constant after emission (unlike the Standard Model Hawking products, which thermalize), and the instantaneous change in the momentum distribution of emitted particles is
\begin{equation}
    \dd{\nH}{p}{t}
    = \nBH \dd{\NH}{p}{t}
    =  \frac{ \nBH }{2\pi} \frac{g_{\rm H} \Gamma_{\rm H}}{e^{p/T_\text{BH}} \pm 1}~~,
\end{equation}
where $\nH$ is the number density of Hawking particles per unit momentum $p$ at cosmic time $t$, $\NH$ is the total number of particles emitted from each black hole, $\gH$ is the number of spin degrees of freedom, and $\Gamma_H$ is the species graybody factor defined in Section~\ref{sec:hawking-radiation}.
Integrating over the black-hole lifetime gives the momentum distribution of all emitted Hawking particles,
\begin{equation}\label{eq:dnHdp-static}
    \dd{\nH}{p} =
     \frac{ \gH \nBH }{2\pi} \int_{0}^{\taubh}\frac{ dt \Gamma_{\rm H}}{e^{p/T_\text{BH}} \pm 1},
\end{equation}
where $\Gamma_{\rm H}$ and $\tbh = (8\pi G \mbh)^{-1}$ depend on time through black-hole evaporation.

Figure~\ref{fig:dndp-static} shows the momentum distribution of a spin-0 particle emitted in a static background.
Compared to a blackbody and a constant-mass black hole, the spectrum from an evaporating black hole is shifted towards higher momenta due to the increasing black-hole temperature as it loses mass.
Most of the high-momentum tail comes from the final $\sim 10\%$ of the lifetime.
The final spectrum differs significantly from the approximation in which the temperature is treated as constant, even though the temperature does not change much during the first $\sim 90\%$ of the black-hole lifetime.

\subsection{General Spacetime Background}\label{sec:mom-dist-general}

We now turn to the general case, in which emitted particles are redshifted during evaporation as the universe expands. After emission, Hawking particles move freely along geodesics, so their momenta are inversely proportional to the scale factor $a$. We therefore define the comoving momentum $\hat p \defeq a p$, which is conserved for Hawking particles after emission.
The instantaneous rate of change in the comoving number density $\hatnH \defeq a^3 \nH$ of Hawking particles per unit comoving momentum is
\begin{equation}
    \dd{\hatnH}{\hat p}{t}
    = \hatnBH \dd{\NH}{\hat p}{t} = 
        \frac{ \gH   \hat n_\text{BH}  }{2\pi } \frac{ \Gamma_{\rm H}/a}{e^{\hat p/(aT_\text{BH} )} \pm 1}~~,
\end{equation}
where $\hatnBH$ is the comoving number density of black holes. Integrating over the black-hole lifetime, we have
\begin{equation}
\label{eq:dnhat/dphat}
    \dd{\hatnH}{\hat p}
    = \frac{\gH \hatnBH}{2\pi} \int_{0}^{\taubh}
    \frac{ dt \Gamma_{\rm H}/a}{e^{\hat p/(aT_\text{BH} )} \pm 1}~~.
\end{equation}
The number of Hawking particles does not change after black holes evaporate, so $\dd{\hatnH}{\hat p}$ is a constant function of $\hat{p}$ thereafter.
In terms of physical momentum, the differential number density after evaporation $(a \geq \aRH)$ is
\begin{equation}\label{eq:dn/dp:main}
\frac{dn_{\rm H}}{dp} =  
\frac{1}{a^2} \frac{d\hat n_H}{d\hat p} \biggr|_{\hat{p} = a p}~~.
\end{equation}

\begin{figure}[t!]
    \centering
    \includegraphics[width=\textwidth]{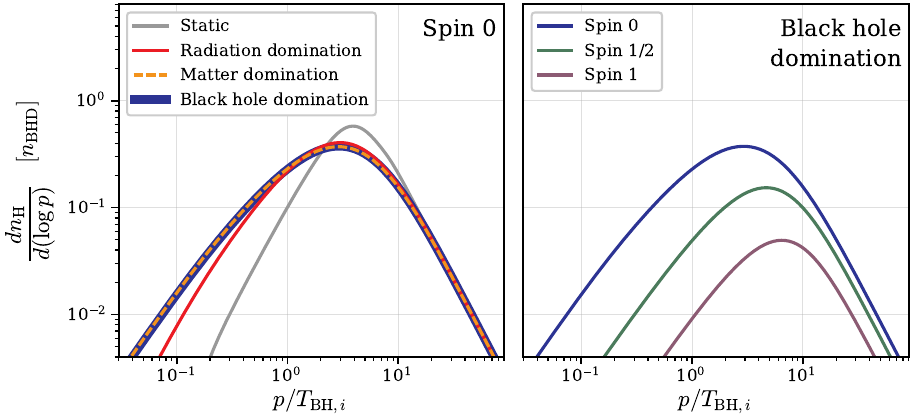}
    \caption{
    Evaporation-time momentum distributions generated by primordial black holes in different cosmological scenarios (left) and for particles with different spin (right). $\nH$ is the number density of emitted particles per degree of freedom as a function of their momenta $p$,
    in units of the total number density $n_{\rm BHD}$ of spin-0 particles in the black-hole-domination scenario. 
    {\bf Left:} Redshifting of particle momenta with the full cosmological background (blue) extends the low-momentum tail compared to the static, non-expanding cosmology (gray).
    {\bf Right:} Particles with higher spin (green, purple) are produced in lower numbers per degree of freedom and skewed towards higher momenta compared to the scalar (blue).
    }\label{fig:dndp-universes-spins}
\end{figure}

In Figure~\ref{fig:dndp-universes-spins}, we show momentum distributions for Hawking particles of various spins evaluated at evaporation time, and we consider the following cosmological scenarios: 
\begin{itemize}
    \item \textbf{Static}, where $a = 1$ throughout the primordial black hole lifetime. In this approximation we neglect the redshifting of particle momenta, as in Section~\ref{sec:mom-dist-static}.
    \item \textbf{Radiation domination}, where $a \propto t^{1/2}$ throughout black-hole evaporation, in which the primordial black holes never contribute significantly to the energy density, as in the left panel of Figure~\ref{fig:density-evol-beta}.
    \item \textbf{Matter domination}, where $a \propto t^{2/3}$ throughout black-hole evaporation, which approximates the black hole-domination scenario.
    \item \textbf{Black-hole domination}, in which black holes dominate the early cosmic energy density and smoothly evaporate into radiation, as
    in the right panel of Figure~\ref{fig:density-evol-beta}. In contrast with Matter Domination as described above, here the accumulating radiation slightly modifies Hubble expansion throughout this era.
\end{itemize}
As Figure~\ref{fig:dndp-universes-spins} shows, the expansion of the universe during evaporation significantly extends the low-momentum tail compared to a static universe, but the high-momentum tail is unaffected because it is produced rapidly at the end of the black-hole lifetime.
At evaporation time, the matter-domination approximation closely reproduces the momentum distribution of black-hole domination because the two cosmologies only differ near the end of the black-hole lifetime, where redshifting effects are not significant.
However, the matter-domination approximation also generates a different reheating temperature via Eq.~\eqref{eq:rhoBH_evap1}, which affects how the momentum distribution is evolved to present day.

The right panel of Figure~\ref{fig:dndp-universes-spins} compares the evaporation-time momentum distribution of Hawking particles with different spins.
Particles with higher spin are produced in lower numbers (per degree of freedom) and skewed towards higher momenta, owing to the spin dependence of the graybody factor \cite{macgibbon_QuarkGluonjetEmission-I-InstantaneousSpectra_1990}.

\subsection{Scaling with Black-Hole Formation Mass}\label{sec:scaling}

In the black-hole domination scenario, the differential number density $d\nH/dp$ of Hawking particles satisfies an exact scaling law with the black-hole formation mass $\mbhi$. We used this scaling property instead of separately solving for the cosmological evolution and momentum distribution for each $\mbhi$.

Let $n_{\rm H, 1}$ and $n_{\rm H, 2}$ denote the number density of Hawking particles that would be generated if primordial black holes had initial mass $\mbhi = M_1$ and $M_2$, respectively, and suppose that the particle spin and degrees of freedom are the same both scenarios. Then at a given scale factor $a$ relative to the present-day (and after evaporation), we have
\begin{equation}\label{eq:dnH-dp-scaling}
    \brac{M_1}{\eta_1^2} \dd{n_{\rm H, 1}}{p} \biggr|_{p=p_1}  \! =
    \brac{M_2}{\eta_2^2} \dd{n_{\rm H, 2}}{p} \biggr|_{p=p_2} ~~,
    \quad
    \frac{p_1}{\eta_1 M_1^{1/2}} = \frac{p_2}{\eta_2 M_2^{1/2}},
\end{equation}
where $\eta_1$ and $\eta_2$ are evaluated in Eq.~\eqref{eq:eta}.
To derive this relation from Eq.~\eqref{eq:dn/dp:main}, we used that $\Gamma_H$ is a function of $GMp$ \cite{page_ParticleEmissionRates-II-MasslessRotating_1976}, that $a / \aRH$ is a function of $t / \taubh$ (see Appendix~\ref{appendix:background-evol-BHdom}), and Eqs.~\eqref{eq:tau}, \eqref{eq:hat-nBH-BH-dom}, and~\eqref{eq:aRH-Mform}. 

\begin{figure}[t!]
    \centering
    \includegraphics[width=\textwidth]{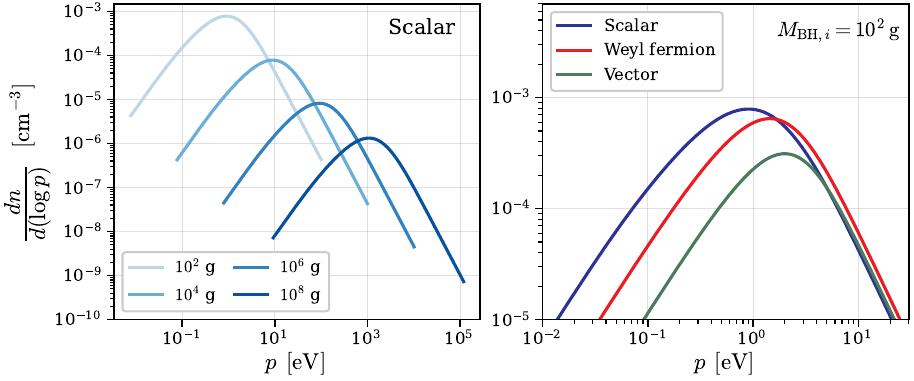}
    \caption{Present-day momentum distributions of Hawking relic particles. \textbf{Left:} 
    Larger primordial black holes leave behind Hawking relics with higher momenta but lower number density. Each line corresponds to a scalar Hawking relic (spin $0$, $g=1$) for a different value of the initial black-hole mass. \textbf{Right:} Comparison of scalar (spin $0$, $g=1$), Weyl fermion (spin $1/2$, $g=2$), and vector (spin $1$, $g=2$) Hawking relics with initial black-hole mass of $\mbhi = \SI{e2}{\gram}$. Each line scales with $\mbhi$ according to Eq.~\eqref{eq:dnH-dp-scaling}.}\label{fig:spin0-dndp-today}
\end{figure}

Figure~\ref{fig:spin0-dndp-today} shows how the present-day momentum distribution varies with the black-hole formation mass and particle spin. Larger black holes leave behind Hawking relics with higher momenta but lower number density. The momenta are higher because larger black holes evaporate later, so Hawking particles are redshifted less between evaporation time and today, with the peak momentum satisfying, by Eq.~\eqref{eq:dnH-dp-scaling},
\begin{equation}\label{eq:p-scaling}
    p \propto \eta \mbhi^{1/2}.
\end{equation}
Integrating Eq.~\eqref{eq:dnH-dp-scaling}, the total number density scales with the initial black-hole mass as
\begin{equation}\label{eq:nH-scaling}
    \nH \propto \eta^3 \mbhi^{-1/2}.
\end{equation}
The number density is smaller for larger formation masses because there are fewer black holes ($\hatnBH \propto \eta^3 \mbhi^{-5/2}$ by Eq.~\eqref{eq:hat-nBH-BH-dom}), even though each black hole produces more particles ($\NH \propto \mbhi^2$). We note that for all formation masses in Figure~\ref{fig:spin0-dndp-today}, the number density of the Hawking relic is much lower than the relic photon number density of $\sim \SI{e2}{\per\cm\cubed}$, but its mean momentum is much higher than the mean photon momentum of $\sim \SI{e-4}{\eV}$. As we will discuss further in the following section, this sets Hawking relics apart from any other hypothetical relic particles that were initially in thermal contact with the Standard Model.

\section{Observing Hawking Relics}\label{sec:observing}

In many established and theoretical cosmological scenarios, populations of stable, weakly-interacting particles are generated in the early universe and survive to the present day as \textit{relic particles}.
A well-known example is that of relic neutrinos, which decoupled from the Standard Model just before BBN. These belong to the category of \textit{(light) thermal relics}, which are particles that decoupled from the Standard Model plasma while relativistic \cite{green_MessengersEarlyuniverse_2019} (we only consider this type of thermal relic, as opposed to heavier relic particles like WIMPs, which were non-relativistic when they decoupled).

The Hawking relics we consider are relativistic when produced, so they share similarities with light thermal relics.
Both Hawking relics and thermal relics can have significant particle velocities during cosmological structure formation, even as they redshift and become non-relativistic.
However, whereas thermal relics have temperatures and number densities comparable to those of CMB photons, Hawking relics are always much hotter and much less numerous than photons.\footnote{This is true for all black hole-masses considered in this paper. Larger masses could produce Hawking relics with softer momenta relative to Standard Model thermal relics, but from Figure~\ref{fig:Tevap-vs-Mform} these masses would not be compatible with BBN.} As we will see, this allows Hawking relics to have much higher masses than light thermal relics while affecting cosmological observables in similar ways.

The cosmological effects of a Hawking relic depend on its mass relative to the Hawking temperature at black-hole formation time, and we distinguish between three qualitatively distinct regimes:
\begin{itemize}
    \item {\bf  Ultra-light Hawking relics} (dark radiation): If a Hawking relic is light enough to remain relativistic over the age of the universe, its primary effect is on the expansion rate through its contribution to the relativistic energy density. In this case, the Hawking relic is indistinguishable from other ultra-light relics (i.e., its effects are described completely by $\Delta N_{\rm eff}$) and the shape of its distribution function does not directly affect any cosmological observables \cite{hooper_DarkRadiationSuperheavy_2019,hooper_HotGravitonsGravitational_2020,lunardini_DiracMajoranaNeutrino_2020,masina_DarkMatterDark_2020,arbey_PrecisionCalculationDark_2021,cheek_EvaporationPrimordialBlack_2022,cheek_RedshiftEffectsParticle_2022}.

    \item{\bf Ultra-heavy Hawking relics} (cold dark matter): If the mass of a relic exceeds the initial Hawking temperature of the black hole, its production is exponentially suppressed until the black hole becomes hotter over the course of its evaporation. Consequently, such heavy relics are only produced appreciably near the end of the black hole's lifetime and their number densities are suppressed compared to those of lighter relics. Furthermore, since heavy relics have low thermal velocities during structure formation, they constitute some or all of the cold dark matter \cite{allahverdi_NonthermalProductionDark_2018,bernal_GravitationalDarkMatter_2021,bernal_SuperradiantProductionHeavy_2022,fujita_BaryonAsymmetryDark_2014,lennon_BlackHoleGenesis_2018,morrison_MelanopogenesisDarkMatter_2019,hooper_DarkRadiationSuperheavy_2019,masina_DarkMatterDark_2020,gondolo_EffectsPrimordialBlack_2020,bernal_DarkMatterTime_2021,cheek_PrimordialBlackHole-I-SolelyHawking_2022,cheek_EvaporationPrimordialBlack_2022,cheek_PrimordialBlackHole-II-Interplay_2022a,cheek_RedshiftEffectsParticle_2022,Haque:2024cdh}.
    
    \item{\bf Warm Hawking relics} (warm/hot dark matter): If a Hawking relic is sufficiently light to be produced relativistically throughout black-hole evaporation, but also sufficiently heavy to become non-relativistic in the later universe, it behaves like warm dark matter and affects the growth of large-scale structure. 
    Due to their impact on structure formation, these relics cannot constitute all of the dark matter \cite{fujita_BaryonAsymmetryDark_2014,morrison_MelanopogenesisDarkMatter_2019,baldes_NoncoldDarkMatter_2020,gondolo_EffectsPrimordialBlack_2020,lennon_BlackHoleGenesis_2018,masina_DarkMatterDark_2020,auffinger_BoundsWarmDark_2021,masina_DarkMatterDark_2021,cheek_PrimordialBlackHole-I-SolelyHawking_2022,cheek_EvaporationPrimordialBlack_2022,cheek_RedshiftEffectsParticle_2022,Haque:2024cdh}.
\end{itemize}

In this section, we review the observational effects of ultra-light Hawking relics and study, for the first time in detail, the effects of warm Hawking relics that constitute only a fraction of the dark matter (also see Ref.~\cite{auffinger_BoundsWarmDark_2021}). 
We assume throughout that the early universe went through a period of black-hole domination as described in Section~\ref{sec:background-evolution}.
To determine the boundary between the ultra-light and warm relic regimes, we use Eq.~\eqref{eq:dn/dp:main} and~\eqref{eq:dnH-dp-scaling} to compute the mean momentum\footnote{$\mean{\pH}$ is approximately proportional to (spin + 1), but we believe this is a happy accident and does not arise for any fundamental reason.}
for each type of relic at time $t$,
\begin{align}\label{eq:mean-pH}
    \mean{\pH}
    = \frac{1}{n_{\rm H}} \int_0^\infty dp \, p \frac{dn_{\rm H}}{dp}
        \,  = \, \frac{ \SI{0.14}{\eV} \, \eta}{a(t)} \left(\frac{\mbhi}{\si{\gram}}\right)^{1/2} \times
    \begin{cases}
        \num{1.0}, & \text{spin $0$}, \\
        \num{1.5}, & \text{spin $1/2$}, \\
        \num{2.0}, & \text{spin $1$},  \\
        \num{3.1}, & \text{spin $2$},
    \end{cases}
\end{align}
where $\eta$ is given in  Eq.~\eqref{eq:eta}. Thus, a Hawking relic is ultra-light (i.e., relativistic today) if
\begin{equation}
    \mH
    \gtrsim \SI{0.1}{\eV} \, \left(\frac{\mbhi}{\si{\gram}}\right)^{1/2},
\end{equation}
whereas it is warm if 
\begin{equation}
    \SI{0.1}{\eV} \, \left(\frac{\mbhi}{\si{\gram}}\right)^{1/2} \lesssim \mH \lesssim \SI{e13}{\GeV} \brac{\rm g}{\mbhi},
\end{equation}
where the upper bound is Eq.~\eqref{eq:m-upperbound}.
For $\mbhi = \SI{e8}{\gram}$, the cutoff between relativistic and non-relativistic Hawking relics is $\sim \SI{1}{\keV}$, which is significantly larger (by a factor of $\sim 10^4$) than the corresponding cutoff for light thermal relics.

\subsection{Effects of Ultra-Light Hawking Relics}

The primary effect of an ultra-light Hawking relic is its contribution to the radiation density. After electron-positron annihilation, its energy density relative to photons is
\begin{align}\label{eq:rhoH-rel}
    \frac{\rho_\text{H}}{\rho_\gamma}
    = \num{1.2e-2} \, \gH \, \eta^4
    \times
    \begin{cases}
        \num{1.0}, & \text{spin $0$}, \\
        \num{0.55}, & \text{spin $1/2$}, \\
        \num{0.23}, & \text{spin $1$},  \\
        \num{0.026}, & \text{spin $2$}.
    \end{cases}
\end{align}
Since $1 \leq \eta^4 \lesssim 2$ for all black-hole formation masses we consider, this ratio only varies by a factor of 2 across the entire range of formation masses.

Additional relativistic degrees of freedom are a common prediction of beyond-Standard-Model physics.
Their effect can be encapsulated by a shift in the effective number of thermal neutrino species~\cite{aghanim_Planck2018Results_VI_2020},
\begin{equation}
    \Delta N_\text{eff} \defeq \frac{8}{7} \left(\frac{11}{4}\right)^{4/3} \frac{\rho_\text{H}}{\rho_\gamma}.
\end{equation}
A positive value of $\Delta \Neff$ implies the presence of more radiation around recombination, and thus a higher expansion rate $H$.
This produces additional damping in the CMB high-$\ell$ tail and shifts the baryon acoustic oscillation (BAO) peaks~\cite{hou_2013_CMB}.
Together, these effects can be observed in the CMB data and disentangled from other cosmological parameters.
Current CMB observations constrain $|\Delta \Neff| < 0.3$~\cite{aghanim_Planck2018Results_VI_2020}, though future observations will be able to measure deviations an order of magnitude smaller~\cite{abazajian2016cmbs4}.
As a comparison, we show in
Table~\ref{tab:Neff-Omega} the $\Delta \Neff$ predicted for several species of Hawking relics (see also Ref.~\cite{hooper_DarkRadiationSuperheavy_2019}, and Refs.~\cite{hooper_HotGravitonsGravitational_2020,arbey_PrecisionCalculationDark_2021} for spinning black holes). All of these $\Delta \Neff$ values are allowed by current CMB constraints, and barring the graviton case can potentially be probed by future CMB observations as they give rise to $\Delta N_{\rm eff}\gtrsim 0.03$.

\begin{table}[t!]
    \centering
    \begin{tabular}{c | c | c | c }
    \hline
    Name & Spin & $\gH$ & $\eta^{-4} \, \Delta \Neff$ \\
    \hline
    Scalar & 0 & 1 & \num{5.1e-2} \\
    Weyl fermion & 1/2 & 2 & \num{5.6e-2} \\
    Dirac fermion & 1/2 & 4 & \num{1.1e-1} \\
    Massless vector$^\dagger$ & 1 & 2 & \num{2.3e-2} \\
    Graviton & 2 & 2 & \num{2.6e-3}  \\
    \hline
    \end{tabular}
    \caption{
    Prediction for the shift $\Delta N_{\rm eff}$ in the effective number of relativistic neutrinos for ultra-light Hawking relics of different spins. The contribution to $\Delta N_{\rm eff}$ is generic for all primordial black-hole masses when corrected by the order-unity $\eta$ factor in Eq.~\eqref{eq:eta}. $^\dagger$A vector with Stueckelberg mass would have $\gH = 3$, increasing $\Delta \Neff$ by a factor of 1.5.}\label{tab:Neff-Omega}
\end{table}

\subsection{Effects of Warm Hawking Relics on Large-Scale Structure}\label{sec:p(k)}

Warm Hawking relics contribute to the matter component of the universe and therefore influence its gravitational clustering. At the present day, the energy density of a warm Hawking relic is given by
\begin{align}\label{eq:OmegaH}
    \OmegaH h^2 \defeq \frac{\rhoH}{\rho_c} h^2
    &= \num{2.0e-6} \gH \,\eta^3 \left(\frac{\mH}{\si{\eV}}\right) \left(\frac{\si{\gram}}{\mbhi}\right)^{1/2}
    \times
    \begin{cases}
        \num{1.00}, & \text{spin $0$}, \\
        \num{0.37}, & \text{spin $1/2$}, \\
        \num{0.11}, & \text{spin $1$},
    \end{cases}
\end{align}
where $\rho_c \defeq 3 H_0^2 / (8 \pi G) $ is the critical density and $h$ is defined by $H_0 = 100 h \, \si{\km\per\second\per\mega\parsec}$.
In principle, the relic's mass could be chosen so that its energy density matches the dark matter density of $\Omega_\text{DM} h^2 \approx 0.12$ \cite{aghanim_Planck2018Results_VI_2020}.
However, its effects on structure formation limit its maximum energy density to only a small fraction of the total dark matter density (which we will calculate in Section~\ref{sec:constraints}).

Warm Hawking relics have the following main effects on cosmological structure formation:
\begin{itemize}
    \item \textbf{Background expansion}: Hawking relics contribute to expansion of the universe through the Friedmann equation, initially redshifting like radiation before becoming non-relativistic and redshifting like pressureless matter. 
    \item \textbf{Free streaming}:
    Hawking relic velocities remain significant throughout cosmic history, which suppresses the matter power spectrum on small scales. Specifically, the thermal velocity of a relic sets a horizon beyond which it does not gravitationally collapse. In Fourier space, this is expressed as a \textit{free-streaming scale} \cite{ali-haimoud_EfficientImplementationMassive_2013}
    \be
        k_{\rm fs} (a) &= a H \left[ 3 \,\Omega_{\rm M}(a) \left< v_{\rm H}^{-2}\right> /2 \right]^{1/2}  ,
    \ee
where $\mean{v_{\rm H}^{-2}}$ is the mean inverse-square relic velocity and $\Omega_{\rm M}$ is the density of all non-relativistic matter, which has the present-day value $h^2 \Omega_{\mathrm{M},0} \approx 0.1434$ \cite{aghanim_Planck2018Results_VI_2020}. For various spins, we find 
        \be
        k_{\rm fs} \approx \SI{e-2}{\per\mega\parsec} \, \frac{a^2}{\eta} \frac{H}{H_0} \brac{\Omega_{\rm M}}{\Omega_{\mathrm{M},0}}^{1/2} \biggl( \frac{m_{\rm H}}{\rm eV} 
        \biggr)\left(\dfrac{\rm g}{\mbhi}\right)^{1/2}
        \times
        \begin{cases}
            \num{1.0}, & \text{spin $0$}, \\
            \num{0.61}, & \text{spin $1/2$}, \\
            \num{0.44}, & \text{spin $1$}.
        \end{cases}
        \label{eq:kfs}
    \ee
      Faster Hawking relics move too quickly to become trapped by feeble gravitational potentials, and therefore do not cluster on  small scales.
    As a consequence, Hawking-relic fluctuations $\delta_{\rm H}$ (where $\delta_i \equiv \rho_i / \mean{\rho_i} - 1$ for any fluid $i$) will be suppressed as $(k/k_{\rm fs})^{-2}$ for modes below the free-streaming scale $(k > k_{\rm fs})$, thereby washing out their small-scale fluctuations.
    
    The suppression of Hawking-relic fluctuations also affects the overall clustering of matter. Matter fluctuations consist of two components, one cold (cold dark matter and baryons) and one warm (Hawking relics):
    \be
    \delta_{\rm m} \equiv f_{\rm cb} \delta_{\rm cb} + f_{\rm H} \delta_{\rm H},~~~ f_i \equiv \rho_i/\rho_\text{m},
    \ee
    where $f_{\rm cb}$ and $f_{\rm H}$ are the matter fractions of cold dark matter+baryons (cb) and Hawking relics (H), respectively.
    Note that even if the Hawking relic is only a subdominant component of all matter ($f_{\rm H} \ll f_{\rm cb}$), its presence can still significantly suppress $\delta_{\rm m}$.
    This is because the absence of $\delta_{\rm H}$ fluctuations for $k>k_{\rm fs}$ yields slower growth for $\delta_{\rm cb}$ fluctuations, which can be understood from the Hawking relic contributing as matter to the Hubble rate $H$, but not to matter clustering ($\delta_{\rm H}\to 0$ on small scales).
    For neutrinos, this ``backreaction'' gives rise to a well-known suppression in the matter power spectrum of (relative) size $-8 f_{\nu}$~\cite{lesgourgues_MassiveNeutrinosCosmology_2006}, where $f_{\nu}$ is the neutrino matter fraction.
    Hawking relics are typically slower than neutrinos, and therefore become non-relativistic earlier and suppress fluctuations for longer durations. Thus, for Hawking relics we expect the corresponding suppression factor  to be  larger \cite{boyarsky_LowerBoundMass_2009} and that the degeree of small scale power suppression (for $k>k_{\rm fs}$) depends on the relic's matter fraction $f_{\rm H}$; power on larger scales will be largely unaffected.
\end{itemize}

To determine the impact of a Hawking relic on the matter power spectrum, we used the Boltzmann solver \texttt{CLASS} \cite{blas_CosmicLinearAnisotropy_2011} and computed the relic distribution functions today using Eq.~\eqref{eq:dn/dp:main}. These distributions serve as the input for a generic non-cold dark matter ({\tt ncdm}) component in \texttt{CLASS}, which self-consistently evolves the evolution of fluctuations in all fluid components, including the Hawking relic.
Beyond adding the Hawking relic, we used parameters from the standard $\Lambda$CDM cosmology as described in Section~\ref{sec:intro}. We then compared the power spectrum with the Hawking relic to the power spectrum generated by \texttt{CLASS} for the base $\Lambda$CDM cosmology.

\begin{figure}[t!]
    \centering
    \includegraphics[width=\textwidth]{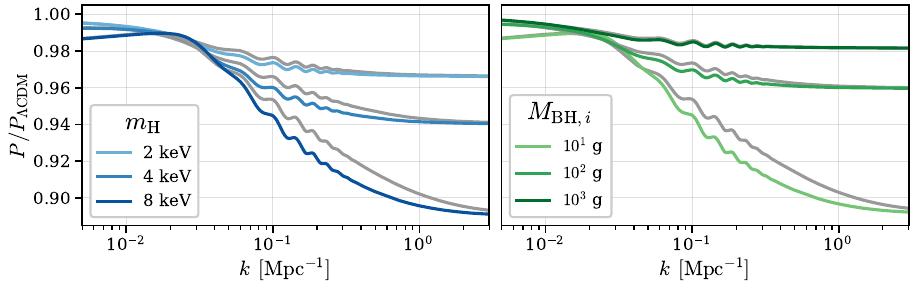}
    \caption{Suppressive effects of scalar Hawking relics on the linear matter power spectrum $P(k)$ compared to the base $\Lambda$CDM (i.e., no relic) prediction.
    {\bf Left:} We show different Hawking particle masses $\mH$ in blue lines, keeping the black-hole formation mass fixed at \SI{e2}{\gram}.
    Gray lines show the closest equivalent thermal relic for each case, which agree at asymptotically large and small $k$ but deviate in the intermediate regime. The wiggles at $k\sim 0.1$ Mpc$^{-1}$ reflect shifts in the BAO peaks~\cite{baumann_PhasesNewPhysics_2017}.
    {\bf Right:} As in the left plot, but varying the black-hole formation mass $\mbhi$ while keeping $\mH = \SI{2.5}{\keV}$ fixed.
    These parameters were chosen such that the deepest suppression is at the boundary of the allowed region of the constraints in Section~\ref{sec:constraints}. The fact that the \SI{8}{\keV} Hawking particle in the left panel (with $\mbhi = \SI{e2}{\gram}$) has a similar effect to the \SI{2.5}{\keV} Hawking particle in the right panel (with $\mbhi = \SI{e1}{\gram}$) is no accident; as shown in Section~\ref{sec:equiv-thermal}, the ratio $\mbhi^{1/2} / \mH$ governs the suppression in the power spectrum.}\label{fig:Pk-mH-Mform}
\end{figure}

Figure~\ref{fig:Pk-mH-Mform} shows the effects of different Hawking relics on the linear matter power spectrum\footnote{We only show results for linear theory, as relics of different masses can change the nonlinear predictions due to their clustering~\cite{loverde_NeutrinoClusteringSpherical_2014,banerjee_SignaturesLightMassive_2022}.}.
Each relic begins to suppress the power spectrum at $k \sim \SI{e-2}{\per\mega\parsec}$ and reaches maximal suppression near its free-streaming scale at $k_{\rm fs} \sim \SI{e0}{\per\mega\parsec}$. The magnitude of suppression is greater for heavier relics because they constitute a larger fraction of the dark matter density, and it is shifted towards smaller scales (larger $k$) as their mean velocities are smaller. The size of the suppression is approximately $-10 f_{\rm H}$, where $f_{\rm H}$ is the relic's fraction of the matter density.
The right panel of Figure~\ref{fig:Pk-mH-Mform} shows how the suppression varies with the black-hole formation mass $\mbhi$: it is larger for lighter primordial black holes because the number density of Hawking relics is larger, and shifts towards smaller scales (larger $k$) because their momenta are smaller (see Section~\ref{sec:scaling}). All of the Hawking relics in Figure~\ref{fig:Pk-mH-Mform} are allowed by current observational constraints and could be detected by measuring the matter power spectrum at $k \gtrsim \SI{e-1}{\per\mega\parsec}$ to within a few per cent, which is potentially achievable by current spectroscopic galaxy surveys like SDSS~\cite{SDSS:2006lmn},  DESI~\cite{desicollaboration_DESIExperimentPart_2016a}, HETDEX~\cite{gebhardt_HobbyEberlyTelescopeDark_2021}, or PFS~\cite{takada_ExtragalacticScienceCosmology_2014}, though the signal requires careful theoretical modeling.

\subsection{Comparison to Light Thermal Relics}
\label{sec:equiv-thermal}

Suppression of the matter power spectrum is not unique to Hawking relics: a light but massive \textit{thermal} relic (LiMR~\cite{xu_CosmologicalConstraintsLight_2022}) would produce a similar---though not identical---effect.
To investigate the differences between these scenarios, we compare each Hawking relic to an ``equivalent'' thermal fermion whose impact on the power spectrum is as close as possible to the Hawking relic. Specifically, we consider a non-interacting fermion with two spin degrees of freedom and distribution function
\begin{equation}\label{eq:fp-fermion}
    f_{\rm th}(p) = \frac{1}{e^{p/T_{\rm th}(t)} + 1}.
\end{equation}
This distribution function generically describes a fermion that was initially in thermal equilibrium with the Standard Model and decoupled while relativistic, in which case its distribution function retains the form Eq.~\eqref{eq:fp-fermion} even as it redshifts and eventually becomes non-relativistic. If the fermion decoupled after electron-positron annihilation, its present-day temperature would simply be the photon temperature of $T_\gamma \approx \SI{2.7}{\kelvin}$, since both the fermion and photon temperatures redshift as $T \propto a^{-1}$. However, if it decoupled prior to $e^+e^-$ annihilation, photons would have received the additional electron entropy but not the fermion, resulting in the photons being hotter than the fermion at the present day. As is well known, the Standard Model neutrinos decoupled just prior to $e^+e^-$ annihilation, so their temperature today is lower than photons, $T_\nu \approx (4/11)^{1/3} T_\gamma \approx \SI{1.9}{\kelvin}$. If the fermion decoupled even earlier, it would have an even lower temperature, down to a minimum possible temperature of around $\SI{0.91}{\kelvin}$ if it decoupled before top quarks became non-relativistic (assuming no additonal entropy transfers from heavy new physics, e.g. supersymmetric particles) \cite{xu_CosmologicalConstraintsLight_2022}.

Our thermal fermion is completely described by its mass $m_{\rm th}$ and its present-day temperature $T_{\rm th}$. In order to make its effects on structure formation as close to the Hawking relic as possible, we choose these parameters so that its energy density and mean velocity match the Hawking relic at the present day, as in \cite{munoz_EfficientComputationGalaxy_2018}. To do this, we first relate its number density and mean particle momentum to $T_{\rm th}$ by
\begin{eqnarray}
    n_{\rm th} &=& \frac{g}{2 \pi^2} \int_0^\infty dp \, p^2 f_{\rm th}(p) = \frac{3 \zeta(3)}{2 \pi^2}  T_{\rm th}^2, \\
    \mean{p_{\rm th}} 
    &=&
    \frac{1}{n_{\rm th}} \frac{g}{2 \pi^2} \int_0^\infty dp \, p^3 f_{\rm th}(p) = \frac{7 \pi^4}{180 \zeta(3)} T_{\rm th},
\end{eqnarray}
from which we calculate its energy density and mean velocity via $\rho_{\rm th} = m_{\rm th} n_{\rm th}$ and $\mean{v_{\rm th}} = \mean{p_{\rm th}} / m_{\rm th}$ because it is nonrelativistic at the present day. Equating these to the Hawking relic using Eq.~\eqref{eq:mean-pH} and~\eqref{eq:OmegaH}, we find that the equivalent thermal relic temperature is 
\begin{align}\label{eq:Tth}
    T_\text{th}
    &= \SI{0.92}{\kelvin} \, \gH^{1/4} \, \eta \times
    \begin{cases}
        \num{1.0}, & \text{spin $0$}, \\
        \num{0.86}, & \text{spin $1/2$}, \\
        \num{0.69}, & \text{spin $1$}, 
    \end{cases}
\end{align}
and the corresponding thermal relic mass is 
\begin{align}\label{eq:mth}
    m_\text{th}
    &= \num{1.8e-3} \, \gH^{1/4} \, \mH \left(\frac{\si{\gram}}{\mbhi}\right)^{1/2} \times
        \begin{cases}
            \num{1.0}, & \text{spin $0$}, \\
            \num{0.57}, & \text{spin $1/2$}, \\
            \num{0.34}, & \text{spin $1$}. 
        \end{cases}
\end{align}
\begin{table}[t]
    \centering
    \begin{tabular}{c | c | c | c | c}
    \hline
    & & & \\[-13pt]
    Name & Spin & $\gH$ & $\displaystyle \left(\frac{\mbhi}{\si{\gram}}\right)^{1/2} \frac{m_\text{th}}{\mH}$ & $\displaystyle \eta^{-1} \, T_\text{th} \,\, [\si{\kelvin}]$ \\
    & & & \\[-13pt]
    \hline
    Scalar & 0 & 1 & \num{1.8e-3} & 0.92 \\
    Weyl fermion & 1/2 & 2 & \num{1.2e-3} & 0.94 \\
    Dirac fermion & 1/2 & 4 & \num{1.4e-3} & 1.1  \\
    Massive vector$^\dagger$ & 1 & 3 & \num{7.9e-4} & 0.84 \\
    \hline
    \end{tabular}
    \caption{Temperature $T_\text{th}$ and mass $m_\text{th}$ of the equivalent thermal fermion for different types of Hawking relic with $g_{\rm H}$ degrees of freedom. $\eta$ is related to the primordial black hole formation mass $\mbhi$ by Eq.~\eqref{eq:eta}, and for $\mbhi \lesssim \SI{e5}{\gram}$ it equals unity, so in this regime all Hawking relics have a universal equivalent temperature today, independent of $\mH$ and $\mbhi$.
    $^\dagger$Here we assume Stueckelberg mass.}\label{tab:mnu-Tnu}
\end{table}

Table~\ref{tab:mnu-Tnu} shows the parameters of the equivalent thermal fermions corresponding to different types of Hawking particles.
Note that the equivalent fermion is much lighter than the Hawking relic itself, by around three orders of magnitude for $\mbhi \sim \SI{1}{\gram}$, and up to seven orders of magnitude for $\mbhi \sim \SI{e8}{\gram}$. This is because thermal relics generically have much higher number densities than Hawking relics (see Section~\ref{sec:scaling}), so their masses must be smaller to match the energy density of Hawking relics. Thus, {\it Hawking relics can be much heavier than thermal relics while having similar cosmological effects}.  
Furthermore, the temperature $T_\text{th}$ of the equivalent fermion is independent of the Hawking relic mass and depends on $\mbhi$ only through $\eta$, which is effectively constant for $\mbhi \lesssim \SI{e5}{\gram}$.
This temperature amounts to a concrete \textit{prediction}: for a given species of Hawking relic (defined by its spin and degrees of freedom), there is a single preferred temperature independent of its mass $\mH$ and the primordial black hole formation mass $\mbhi$ over a wide range of parameter space. Interestingly, owing to the non-thermal origin of Hawking relics, in some cases this preferred temperature is below \SI{0.91}{\kelvin}, the lowest value predicted for a thermal fermion. If the signature of a thermal fermion is detected at one of these preferred temperatures, further investigation is warranted to determine whether it has a Hawking-evaporation origin.

\begin{figure}[htb]
    \centering
    \includegraphics[width=\textwidth]{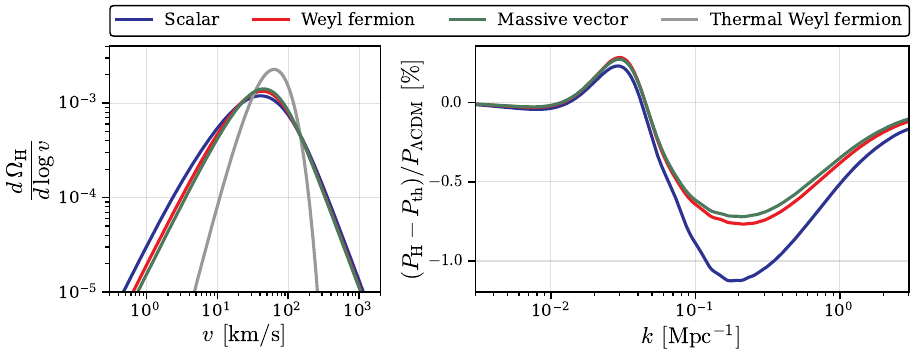} 
    \caption{
    \textbf{Left}: Present-day velocity distributions of warm Hawking relics compared to a thermal Weyl fermion, all with the same total energy density and mean particle velocity.
    The scalar Hawking relic has mass $\mH = \SI{1.6}{\MeV}$ with $\mbhi = \SI{5.6e6}{\gram}$, the fermion $\mH = \SI{10}{\keV}$ with $\mbhi = \SI{e2}{\gram}$, and the massive vector $\mH = \SI{9.0}{\MeV}$ with $\mbhi = \SI{3.5e7}{\gram}$.
    The thermal fermion has mass $m_\mathrm{th} = \SI{1.21}{\eV}$ and temperature $T_\mathrm{th} = \SI{0.94}{\kelvin}$, as determined from Table~\ref{tab:mnu-Tnu}.
    \textbf{Right}: Percentage change in the matter power spectrum $P(k)$ for each Hawking relic compared to the thermal fermion.
    Hawking relics produce larger suppression than the thermal fermion at $k \gtrsim \SI{0.04}{\per\mega\parsec}$,
    with the greatest difference at $k \sim \SI{0.2}{\per\mega\parsec}$,
    though the difference between the Hawking and thermal relics is $\lesssim 1\%$ (whereas the difference to $\Lambda$CDM with no relic is $\sim 7\%$). The scalar is the most distinguishable from the thermal fermion due to its wider velocity distribution.
    }\label{fig:spin0-drhodv-Pk-spins}
\end{figure}

Returning to the cosmological effects of Hawking relics, we show in Figure~\ref{fig:Pk-mH-Mform} the suppression in the matter power spectrum of the equivalent fermion alongside each Hawking relic. The two relics affect the power spectrum in broadly similar ways. For example, an \SI{8}{\keV} scalar Hawking relic with $\mbhi = \SI{e2}{\gram}$ has a similar effect to its equivalent \SI{1.4}{\eV} thermal fermion with temperature \SI{0.92}{\kelvin}, both resulting in around $11\%$ suppression in the linear power spectrum at large $k$. However, their effects are not identical: the \SI{8}{\keV} Hawking relic produces approximately $1 \%$ more suppression at observable scales ($k = \SI{e-1}{\per\mega\parsec}$) relative to its equivalent thermal fermion. The subtle differences between the two types of relics, which makes them potentially distinguishable by high precision measurements, arise from their different velocity distributions.
As we show in Figure~\ref{fig:spin0-drhodv-Pk-spins} (left panel), the velocity distributions of warm Hawking relics are much wider compared to thermal fermions due to the combined effects of the black-hole temperature increasing and the universe expanding during evaporation.
Of the three types of Hawking relics shown in Figure~\ref{fig:spin0-drhodv-Pk-spins}, scalars have the widest velocity distributions and are thus the most distinguishable from thermal relics.

\begin{figure}[t!]
    \centering
    \includegraphics[width=\textwidth]{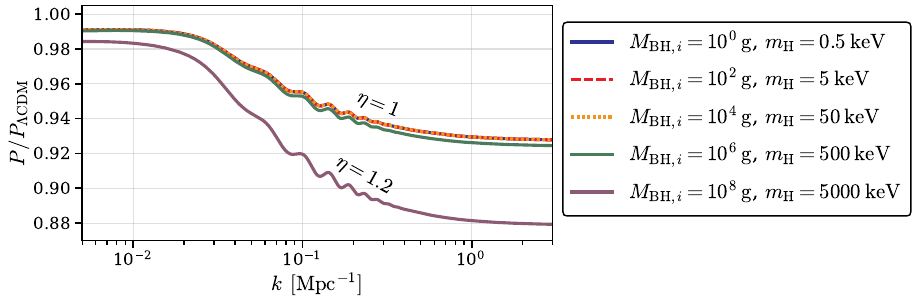}
    \caption{Same as Figure~\ref{fig:Pk-mH-Mform}, but for scalar Hawking relics with mass $\mH$ produced from primordial black holes with formation mass $\mbhi$, such that the ratio $\mH / \mbhi^{1/2}$ is held constant.
    For $\mbhi \lesssim \SI{e5}{\gram}$, in which case $\eta = 1$, two Hawking particles with the same $\mH / \mbhi^{1/2}$ have identical energy density per unit velocity, and this means that their effects on the power spectrum are identical.}\label{fig:mH-Mform-degeneracy}
\end{figure}

While it is possible to identify a Hawking relic in the matter power spectrum, measuring its properties will be more difficult.
Figure~\ref{fig:spin0-drhodv-Pk-spins} shows that the signatures of a Weyl fermion and a massive vector are nearly identical, making the prospect of distinguishing the two cases extremely challenging.
Moreover, the mass of a Hawking relic cannot always be identified from the power spectrum due to a degeneracy with the black-hole formation mass.
The velocity of a non-relativistic Hawking relic is proportional to $\eta \mbhi^{1/2} / \mH$ by Eq.~\eqref{eq:p-scaling}, while its energy density is proportional to $\eta^3 \mH / \mbhi^{1/2}$ by Eq.~\eqref{eq:OmegaH}. Therefore, two Hawking particles have identical energy density per unit velocity if $\mH / \mbhi^{1/2}$ is held constant, provided $\mbhi \lesssim \SI{e5}{\gram}$ (so that $\eta = 1$), and this means that their effects on the power spectrum are identical. We show this in Figure~\ref{fig:mH-Mform-degeneracy}.

\subsection{New Constraints on Hawking Relics}\label{sec:constraints}

Since the effect of a Hawking relic on the power spectrum is well approximated by an equivalent thermal fermion, we can translate existing constraints on thermal fermions into new limits for Hawking relics. To calculate these constraints, for each type of Hawking particle and for each black-hole formation mass $\mbhi$, we calculated the temperature $T_\text{th}$ of the equivalent thermal fermion using Table~\ref{tab:mnu-Tnu}. We then found the maximum allowed thermal fermion mass at this temperature allowed by the 95\% confidence limit from a joint analysis of CMB, galaxy clustering, and weak-lensing data presented in Ref.~\cite{xu_CosmologicalConstraintsLight_2022}, and translated it into the maximum Hawking relic mass, again using Table~\ref{tab:mnu-Tnu}. This does not give an exact bound because Hawking relics have slightly different effects on the power spectrum compared to thermal relics. Nonetheless, we expect this approximation to suffice for our current data, as Hawking relics show very similar suppression compared to equivalent thermal fermions at very low and high $k$, with only a modest relative increase in suppression at intermediate scales (see Figure~\ref{fig:Pk-mH-Mform}).

\begin{figure}[t!]
    \centering 
    \includegraphics[width=0.75\textwidth]{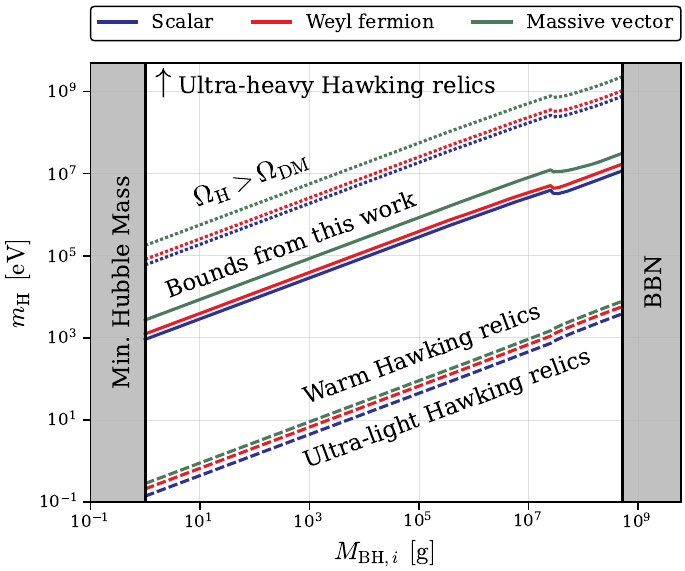}
    \caption{Constraints on warm Hawking relics from primordial black-hole domination.
    The regions above the solid lines are ruled out for warm Hawking relics of mass $m_{\rm H}$ produced by primordial black holes of mass $\mbhi$.
    For reference, dotted lines show where the abundance of the Hawking relic equals all of dark matter and dashed lines show the mass below which the relic would be relativistic today. Our bounds do not constrain ultra-heavy Hawking relics, which can be sufficiently slow to mimic the cosmological clustering of cold dark matter. Furthermore, such relics can evade overclosure constraints and even account for  all of dark matter in the black-hole domination scenario  if $\mH \gtrsim \SI{e18}{\eV}$ \cite{cheek_PrimordialBlackHole-I-SolelyHawking_2022}.
    }\label{fig:constraints}
\end{figure}

Figure~\ref{fig:constraints} shows our constraints on warm Hawking relics in the $\mbhi$--$\mH$ plane. For $\mbhi \lesssim \SI{e5}{\gram}$, all Hawking relics of a given type have the same equivalent fermion temperature, so the constraint is equivalent to an upper bound on their energy density and the maximum Hawking relic mass is exactly proportional to $\mbhi^{1/2}$ (see Eq.~\eqref{eq:OmegaH} with $\eta = 1$). 
For $\mbhi \gtrsim \SI{e5}{\gram}$, the equivalent fermion temperature increases with $\eta(\mbhi)$ and this simple scaling rule no longer holds.
We rule out the possibility that a warm Hawking relic is all the dark matter (indicated by dotted lines in Figure~\ref{fig:constraints}), in agreement with prior studies \cite{fujita_BaryonAsymmetryDark_2014,morrison_MelanopogenesisDarkMatter_2019,baldes_NoncoldDarkMatter_2020,gondolo_EffectsPrimordialBlack_2020,lennon_BlackHoleGenesis_2018,masina_DarkMatterDark_2020,auffinger_BoundsWarmDark_2021,masina_DarkMatterDark_2021,cheek_PrimordialBlackHole-I-SolelyHawking_2022,cheek_EvaporationPrimordialBlack_2022,cheek_RedshiftEffectsParticle_2022,Haque:2024cdh}.
However, there is still a significant region of allowed parameter space for warm Hawking relics. The largest allowed particle mass in Figure~\ref{fig:constraints} is $\sim \SI{50}{\MeV}$ and corresponds to a vector Hawking relic. This is seven orders of magnitude larger than the maximum allowed mass of a thermal vector \cite{xu_CosmologicalConstraintsLight_2022}.
Of the three Hawking relics in Figure~\ref{fig:constraints}, the maximum allowed energy density is $\OmegaH h^2 \sim \num{2e-3}$ for a Weyl fermion, in which case it would constitute $\sim 2\%$ of the dark matter. In fact, a percent-level bound on the total energy density in the form of warm Hawking relics holds even if we allow multiple Hawking relic particles with similar masses. This is because each Hawking relic degree of freedom increases the temperature of the equivalent thermal fermion by Eq.~\eqref{eq:Tth}, reducing its maximum allowed mass.

\begin{figure}[t!]
    \centering 
    \includegraphics[width=\textwidth]{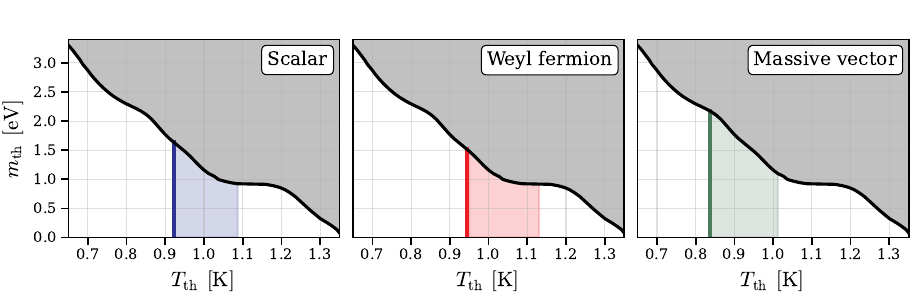}
    \caption{Regions of thermal fermion parameter space corresponding to different Hawking relics. The shaded gray regions are ruled out by Ref.~\cite{xu_CosmologicalConstraintsLight_2022}. 
    The temperature $T_{\rm th}$ of the thermal fermion corresponding to a given black-hole formation mass is given in Table~\ref{tab:mnu-Tnu}. The thermal fermion can have any mass $m_{\rm th}$ below than the constraint line, which translates to a Hawking relic mass $\mH$ by Table~\ref{tab:mnu-Tnu}.
    The solid vertical lines mark the preferred temperatures for each type of Hawking relic; these temperatures are universal for $\mbhi \lesssim \SI{e5}{\gram}$ (i.e. the same for all Hawking relic masses $\mH$ and black-hole formation masses $\mbhi$). The colored shaded regions mark the extended parameter space for heavier black-hole formation masses; these regions are allowed within our constraints and correspond to Hawking relics from primordial black holes with masses $\mbhi > \SI{e5}{\gram}$. 
    }\label{fig:equiv-fermion-spaces}
\end{figure}

Figure~\ref{fig:equiv-fermion-spaces} shows the region of the thermal fermion parameter space that is covered by different Hawking relics, which is another key result of this paper. The vertical solid lines mark the preferred temperatures that apply to all black-hole formation masses below $\sim \SI{e5}{\gram}$ (see Table~\ref{tab:mnu-Tnu}). The shaded regions mark the extended parameter space for heavier black-hole formation masses where the equivalent fermion temperature increases with $\eta(\mbhi)$.
These regions of parameter space are particularly relevant to future cosmological searches for light thermal relics. If a thermal relic is detected in one of these regions (especially at one of the preferred temperatures), further investigation of its momentum distribution will be warranted to determine whether it is a thermal or a Hawking relic.

Our results in this section apply if the early universe underwent a period of black-hole domination.
This is a fairly generic consequence of ultra-light primordial black holes unless their initial abundance $\beta$ was very small (see Figure~\ref{fig:Mform-beta-region}), and in this regime nearly all observables are independent of $\beta$, making it a particularly interesting case to study.
However, if the initial abundance was sufficiently small (e.g. $\beta \lesssim \num{e-10}$ for \SI{e4}{\gram} black holes), black holes would never have dominated the energy density, and the constraints on Hawking relics would depend on $\beta$ and $\mbhi$. For example, as $\beta$ decreases we expect heavier Hawking relics would be allowed because their energy densities would be smaller \cite{cheek_PrimordialBlackHole-I-SolelyHawking_2022}.
But heavier Hawking relics are colder, so their suppressive effects on structure formation are weaker, which means that Hawking relics in the very small $\beta$ regime cannot be tightly constrained by current cosmological data.
Similarly,  our constraints do not apply to ultra-massive Hawking relics that satisfy $m_{\rm H} \gtrsim \SI{e13}{\GeV} (\si{\gram} / \mbhi)$. These relics can avoid over-closing the universe despite their large masses because their production is suppressed until the end of the black-hole lifetime. They are sufficiently massive to be cold today, and so are similarly not tightly constrained by power-spectrum measurements.
Finally, here we have focused on monochromatic primordial black hole mass distributions.
For extended mass distributions, the momentum distribution of Hawking relics is expected to be even wider. Based on our findings in Section~\ref{sec:equiv-thermal}, this would likely increase the distinguishability of Hawking relics from thermally-produced relics. However, properly characterizing these differences would also require a dedicated Hawking-relic search, which is beyond the scope of this work, as the equivalent thermal-relic scenario may not be sufficiently similar to serve as a proxy for a much broader Hawking relic velocity distribution.

\section{Summary and conclusions}\label{sec:conclusion}

Ultra-light primordial black holes ($\mbhi \lesssim \SI{e8}{\gram}$) may have formed in the early universe. Their abundance is only weakly constrained by observations, since their small masses would ensure they evaporated prior to our earliest measurements.
Even if the initial abundance was fairly small, the universe would have gone through an era of black-hole domination before they evaporated, in which case the hot plasma that seeded nucleosynthesis was reheated entirely from evaporating primordial black holes.
Since the Standard-Model particles emitted by Hawking radiation would have thermalized, primordial black hole evaporation produces almost the same outcome as standard hot-big-bang reheating (with some caveats \cite[e.g.,][]{barrow_BaryogenesisExtendedInflation_1991a,inomata_GravitationalWaveProduction_2020}). In other words, primordial black holes may have formed and disappeared almost without a trace.
However, any dark-sector particles in  nature would also be produced by Hawking radiation, populating the early universe and leaving potentially observable signatures of primordial black hole evaporation.

In this paper, we investigated Hawking relics: stable, non-interacting dark-sector particles produced by evaporating primordial black holes in the early universe.
Unlike thermally-produced relics, Hawking relics are gravitational in origin, giving them a distinctive momentum distribution that uniquely influences cosmological observables.
We computed the momentum distribution of Hawking relics in the late universe, taking into account the increasing black-hole temperature and the redshifting of particle momenta during evaporation, and used this to investigate their observable effects.
Our work differs from previous studies that focused on massless relics \cite{hooper_DarkRadiationSuperheavy_2019,hooper_HotGravitonsGravitational_2020,lunardini_DiracMajoranaNeutrino_2020,masina_DarkMatterDark_2020,arbey_PrecisionCalculationDark_2021,cheek_EvaporationPrimordialBlack_2022,cheek_RedshiftEffectsParticle_2022} or massive relics that comprise \textit{all} the dark matter \cite{fujita_BaryonAsymmetryDark_2014,allahverdi_NonthermalProductionDark_2018,lennon_BlackHoleGenesis_2018,morrison_MelanopogenesisDarkMatter_2019,baldes_NoncoldDarkMatter_2020,gondolo_EffectsPrimordialBlack_2020,masina_DarkMatterDark_2020,auffinger_BoundsWarmDark_2021,masina_DarkMatterDark_2021,cheek_PrimordialBlackHole-I-SolelyHawking_2022,cheek_EvaporationPrimordialBlack_2022,cheek_RedshiftEffectsParticle_2022,Haque:2024cdh} in that we considered warm Hawking relics that account for a sub-dominant fraction of dark matter ($\lesssim 2\%$). Our relics occupy a different region of parameter space (i.e., smaller particle masses and larger initial primordial black holes abundances) than hypothetical relics that could account for all the dark matter. The warm relics we considered present a potential target for detection by current and future cosmological measurements of matter density fluctuations.

We computed signatures of warm Hawking relics 
by calculating their distribution functions, adding them to the standard cosmological model, and evolving the linear matter power spectrum. We found that Hawking relics suppress the matter power spectrum for wavenumbers $k \gtrsim \SI{e-2}{\per\mega\parsec}$, with the amount of suppression greater for heavier relic particles and for lighter primordial black holes. This suppression could be detected by current and future cosmological observations, with future surveys expected to improve uncertainties in the suppression amplitude by a factor of two \cite{deporzio_FindingEVscaleLight_2021}. We investigated whether Hawking relics can be uniquely identified by comparing the signature of each Hawking relic to an ``equivalent'' thermally-produced fermion. 
We found that while Hawking relics and thermal relics have very similar effects on the matter power spectrum, they may be distinguishable due to the broader velocity distributions of Hawking relics. However, differentiating between the two would require highly precise measurements of the power spectrum at $k \gtrsim \SI{e-2}{\per\mega\parsec}$.

Our constraints on warm Hawking relics are summarized in Figure~\ref{fig:constraints}.  To calculate these constraints, we approximated each Hawking relic by an equivalent thermal fermion, which produces similar effects on the matter power spectrum.
We found that warm Hawking relics are ruled out by current cosmological data over a broad range of parameter space (including variations of the primordial black holes formation mass $\mbhi$).
This includes the scenario in which a warm Hawking relic constitutes all of dark matter, in agreement with prior studies.
However, there is still a significant region of allowed parameter space, including relics of mass up to $\SI{50}{\MeV}$, which is seven orders of magnitude higher than the maximum allowed mass of a corresponding thermally-produced relic~\cite{xu_CosmologicalConstraintsLight_2022}.
We found that warm Hawking relics can constitute approximately $2\%$ of the dark matter.
Furthermore, in Figure~\ref{fig:equiv-fermion-spaces} we identified three regions in the equivalent-relic parameter space ($m_\text{th}$--$T_\text{th}$) where Hawking relics are expected to appear.
A prospective detection of a thermal relic in one of these regions could indicate that it is, in fact, a Hawking relic. Such a discovery would warrant further investigation to determine whether its momentum distribution corresponds to a thermal or Hawking relic.

In summary, Hawking relics from primordial black holes offer the enticing possibility of probing the earliest moments of the universe.
In this paper, we modeled their evolution throughout cosmic history, from their initial emission as Hawking radiation to their influence on the present-day structure of the universe.
While Hawking relics suppress structure formation in a manner similar to thermally-produced relics, such as massive neutrinos, we found that they leave distinct signatures on the matter power spectrum, allowing them to be identified through measurements of matter density fluctuations.
The discovery of a Hawking relic would open a window to the thermal state of the universe prior to BBN, giving us our earliest glimpse thus far into the the first moments of the universe.
This would not only be important for early-universe cosmology, but it would also open a new frontier of particle physics beyond the Standard Model and give the first observational evidence for Hawking radiation, black-hole evaporation, and primordial black holes.

\acknowledgments

The authors thank Daniel Eisenstein, Dan Hooper, and Linda Xu for insightful discussions.
CJS acknowledges support from the Quad Fellowship.
JBM acknowledges support from the National Science Foundation (NSF) under Grant No.~2307354, and thanks KITP for their hospitality while finishing this paper, which is supported in part by grant NSF PHY-2309135. Fermilab is operated by the Fermi Research Alliance, LLC under Contract DE-AC02-07CH11359 with the U.S. Department of Energy. This material is based partly on support from the Kavli Institute for Cosmological Physics at the University of Chicago through an endowment from the Kavli Foundation and its founder Fred Kavli. 
This work was initiated at the Aspen Center for Physics, which is supported by National Science Foundation grant PHY-2210452.

\appendix

\section{Does Hawking Radiation Thermalize?}\label{appendix:thermalization}

When primordial black holes evaporate at $t = \taubh$, they emit Standard Model (SM) particles at high energies of order $E \sim \tbh$, whose 
cross sections naively scale as $\sigma \sim \alpha^2/\tbh^2$, which might not suffice for thermalizing the Hawking radiation in an expanding, primordial-black-hole dominated universe. However, upon emission, Standard-Model particles can efficiently shower 
massless radiation and redistribute the energy of the Hawking emission to make a larger population of lower energy particles, which subsequently thermalizes in a Hubble time. Here we demonstrate that the Hawking emission does indeed thermalize. 

\subsection*{Naive Estimate}
At face value, thermalization seems like a significant challenge in a primordial-black-hole dominated universe. Upon complete evaporation, each black hole emits approximately  $\sim N_{\rm SM}$ total Standard-Model particles, where 
\be
N_{\rm SM} \sim \brac{\mpl^2}{24\pi} \int_{\tbh}^{\mpl} \frac{d\tbh^\prime}{\tbh^{\prime 3}} \sim 
\frac{1}{48\pi} \brac{\mpl}{\tbh}^2 = 
\frac{4\pi}{3} \brac{\mbh}{\mpl}^2 
\ee  
where we have neglected subdominant contributions from possible other BSM species. The number density of Standard-Model particles at the time of evaporation is 
\be
n_{\rm SM}(\taubh) = N_{\rm SM} \nbh(\taubh)  =N_{\rm SM} \frac{ \rho_{\rm BH}(\taubh) }{\mbh} = 
 \frac{3 N_{\rm SM} \mpl^2 H(\taubh)^2 }{8 \pi \mbh}  = \frac{1}{2} H(\taubh)^2 \mbh,~~
\ee
where we have used the Friedmann equation in black-hole domination. The naive scattering cross section for two color charged particles radiated from the black hole is $\sigma_{\rm SM} \sim \alpha_s^2/\tbh^2$, where $\alpha_s$ is the strong coupling, so the 
scattering rate of Hawking radiation is 
\be
\Gamma_{\rm SM} = n_{\rm SM} \sigma_{\rm SM} =   \frac{32 \pi^2  \alpha_s^2   \mbh^3  }{ \mpl^4}  H(\taubh)^2 ~,
\ee
where $H = 2/(3\taubh)$ in black-hole domination,
so we can substitute for one factor of Hubble to write the rate as 
\be
\Gamma_{\rm SM} =   \frac{64 \pi^2  \alpha_s^2   \mbh^3 }{3  \taubh \mpl^4} H(\taubh) = 
\brac{ \pi^2 \alpha_s^2 g_{\star, H}}{480} H(\taubh)  \sim {\cal O}(H),
\ee
so the scattering rate is on the same order as the Hubble rate and  it is not obvious that thermalization occurs based on this naive argument. Furthermore, since the cross section here did not account for efficient momentum transfer between scattering particles and we omitted order unity factors in the denominator,  it would appear that thermalization should not happen as these effects would only suppress the rate beyond the above estimate.

\subsection*{Including Hadronization}
A proper treatment of thermalization must account for QCD and QED showering effects that enlarge the number of Standard-Model
particles after Hawking evaporation and reduce their energies, thereby greatly enhancing their scattering rates. 
Since most Standard-Model particles are QCD charged, their dynamics will dominate the physics of thermalization and we will neglect QED effects in this estimate. Upon emission from a primordial black hole on a timescale of order $t_{\rm qcd} \sim \Lambda_{\rm qcd}^{-1} \sim 10^{-24}$ s, a quark or gluon showers QCD radiation which yields an average multiplicity of partons 
\be
N_{\rm part} \approx \frac{C_F \alpha_s}{\pi} \log^2\brac{s}{\Lambda_{\rm qcd}^2} ,
\ee
where $s$ is the energy of the jet, $C_F = 3/2$ for an $SU(3)$ gauge theory, 
and $\Lambda_{\rm qcd} \approx 200$ MeV is the scale of QCD confinement. 
 In our scenario $s \sim \tbh^2$ 
so we have 
\be
N_{\rm part} \approx \frac{C_F \alpha_s}{\pi} \log^2\brac{\mpl^4}{64 \pi^2 \mbh^2 \Lambda_{ \rm qcd}^2 }  \approx 120~,
\ee
where, in the last step, we took $\alpha_s = 0.1$ and $\mbh = 10^6$ g. To assess the impact of hadronization, we consider
two regimes below.

\subsubsection*{Fast Hadronization \texorpdfstring{$(\taubh \gg t_{\rm qcd})$}{}}
In this regime, we need $\mbh \gg 10^2$ g so the black-hole lifetime is long compared to the timescale of 
QCD hadronization, and the latter occurs within one Hubble time at the end of black-hole domination. The key differences relative to the naive calculation above are 1) that the number density is enhanced $n_{\rm SM} \to N_{\rm part} n_{\rm SM}$ and 2) that the cross section  is short range and set by the confinement scale $\sigma_{\rm SM} \sim 1/\Lambda_{\rm qcd}^2$ instead of the black-hole temperature. With these modifications, 
the scattering rate now satisfies 
\be
\label{eq:fast-hadronization}
\Gamma_{\rm SM}(\taubh) = N_{\rm part} n_{\rm SM} \sigma_{\rm SM} =  \frac{ N_{\rm part} \mbh  H(\taubh)^2}{2 \Lambda_{\rm qcd}^2 }
\sim 10^{23} H \brac{\rm g}{\mbh}^2
\ee
where we have used $H(\taubh) = 2/(3\taubh)$ at evaporation. Clearly, after hadronization the scattering rate is fast compared to Hubble expansion in the  $\taubh \gg t_{\rm qcd}$ regime.

\subsubsection*{Slow Hadronization \texorpdfstring{$(\taubh \ll t_{\rm qcd})$}{}}

In the slow hadronization regime ($\mbh \lesssim 10^2$), QCD charged particles shower partons at evaporation and this process continues even after evaporation when the universe becomes radiation dominated, so we evaluate the rate at the hadronization time instead of the evaporation time. As in the fast hadronization regime above, here we have the same expected value of $N_{\rm part}$ after hadronization completes, but the Standard-Model scattering rate will be parametrically different as the number density of Standard-Model particles will be additionally redshifted by Hubble expansion, where
\be
n_{\rm SM}(t_{\rm qcd}) =  n_{\rm SM}(\taubh)  \left[\frac{a(\taubh)}{a(t_{\rm qcd})} \right]^3 =  \frac{1}{2} H^2(\taubh) \mbh \brac{\taubh}{t_{\rm qcd}}^{3/2},
\ee
and we have used the fact that $a \propto t^{1/2}$ in radiation domination after evaporation. The extra redshift factor introduces an additional suppression relative to fast hadronization regime studied above, where 
\be
\brac{\taubh}{t_{\rm qcd}}^{3/2} =  \brac{10240 \,\Lambda_{\rm qcd}\, \mbh^3   }{ g_{\star, H}  \mpl^4 }^{3/2} \approx  2 \times 10^{-6} \brac{\mbh}{\rm g}^{9/2},
\ee
which does not compensate for the large scattering rate relative to Hubble in Eq. \ref{eq:fast-hadronization} for any viable choice of $\mbh > 10$ g compatible with black-hole domination in a causal sub-Horizon formation mechanism. Note also that
$\taubh(10 \rm g) \sim 10^{-23}$ s, which is numerically similar to $t_{\rm qcd} \sim 10^{-24}$ s. We thus conclude that Hawking radiation thermalizes in a Hubble time for the full range of masses compatible with an epoch of black-hole domination.

\section{Graybody Factors and Instantaneous Emission Spectrum}\label{appendix:graybody}

\begin{figure}[t!]
    \centering
    \includegraphics[width=\textwidth]{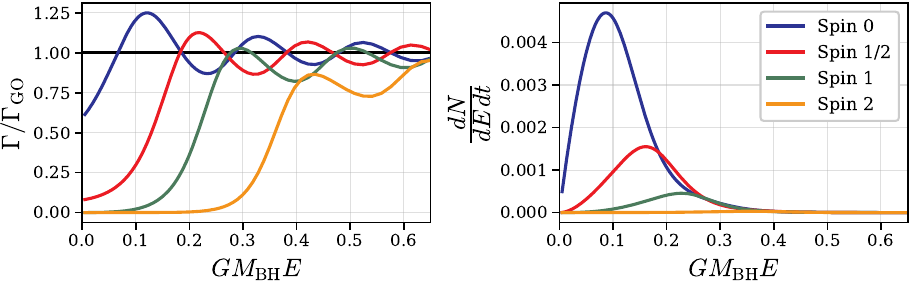}
    \caption{The graybody factor suppresses the emission of low-energy particles, with the amount of suppression greater for particles with higher spins. \textbf{Left:} Graybody factor $\Gamma$ relative to the geometric optics limit $\Gamma_{\rm GO} = 27 (G \mbh E)^2$, where $\Gamma \to \Gamma_{\rm GO}$ as $E \to \infty$. \textbf{Right:} Instantaneous emission spectrum of Hawking radiation per degree of freedom for particles with different spins, as given by Eq.~\eqref{eq:dNdEdt-Gamma}. This plot assumes relativistic emission (i.e., $m \ll T_{\rm BH}$, where $m$ is the particle mass).}\label{fig:graybody}
\end{figure}

The graybody factor $\Gamma_j(E, \mbh)$ accounts for dynamical interactions between the black hole and the field of particles emitted by Hawking radiation.
As a result of these interactions, the black hole will absorb waves for some angular modes and energies, and amplify them for others.
The energy dependence of the graybody factor captures the difference between emission from Hawking radiation and a black body.
It can be calculated using perturbation theory on a dynamical black-hole background \cite{teukolsky_PerturbationsRotatingBlack_1973,teukolsky_PerturbationsRotatingBlack_1974}. 
At high particle energies ($E \gg T_\text{BH}$), the graybody factor approaches the \textit{geometrical-optics limit} $\Gamma_{j} \to \Gamma_{\rm GO} \defeq 27 (G \mbh E)^2$, whereas for low energies it approaches a spin-dependent limit \cite{page_ParticleEmissionRates-I-MasslessNonrotating_1976,macgibbon_QuarkGluonjetEmission-I-InstantaneousSpectra_1990}.
Figure~\ref{fig:graybody} shows the graybody factors for particles of different spins, and the resulting emission spectrum from Hawking radiation. The graybody factor significantly suppresses the emission of low-energy particles relative to the geometric optics limit, with the amount of suppression greater for particles with higher spins.
Throughout this paper, we use the graybody factors tabulated by Ref.~\cite{arbey_BlackHawkV2Public_2019}.

\section{Spacetime Evolution in Black-Hole Domination}\label{appendix:background-evol-BHdom}

\begin{figure}[t!]
    \centering
    \includegraphics[width=\textwidth]{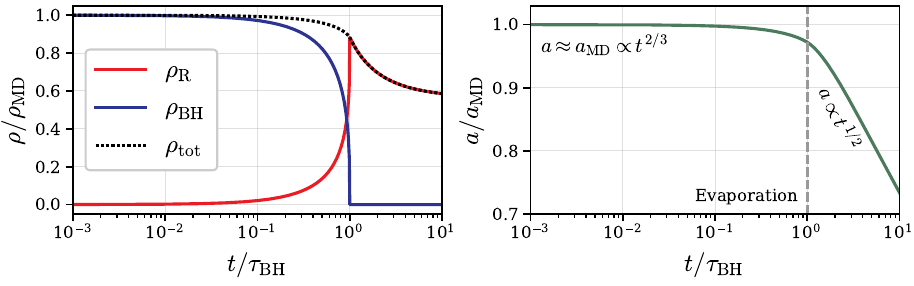}
    \caption{Universal spacetime evolution in black-hole domination.
    {\bf Left:} evolution of the component energy densities relative to a matter-dominated (MD) universe. {\bf Right:} evolution of the scale factor relative to a matter-dominated universe, normalized so that they are equal at early times.}\label{fig:density-evol-universal}
\end{figure}

After primordial black holes form, the universe evolves according to the fluid equations in Eq.~\eqref{eq:fluid}, with black holes losing mass according to Eq.~\eqref{eq:M(t)-approx}.
If we begin our evolution at a time $t$ long after black-hole domination begins, but long before Hawking radiation becomes significant, we can take the initial conditions to be
\begin{equation}\label{eq:BHdom-ICs}
    \rhoBH(t) = \frac{3}{8\pi G} \left(\frac{2}{3t}\right)^2 ~, ~~~~ \rhoR(t) = 0. 
\end{equation}
The value of the initial time $t$ does not matter as long as these conditions are satisfied; we use $t = \num{e-5} \, \taubh$.

The fluid equations can be written in a universal form, independent of the black-hole mass $\mbhi$ and emission rate $\kappa$, by defining
\be
\tilde t \equiv \frac{t}{\taubh}~~,~~
\tilde{M} \equiv \frac{M_{{\rm BH}}}{M_{{\rm BH},i}} ~~,~~\tilde{H} \equiv \taubh H~~,~~ \tilde{\rho} \equiv G \taubh^2 \rho ~,
\ee 
where $\taubh$ is defined in Eq.~\eqref{eq:tau}. In terms of these quantities, the fluid equations do not explicitly depend on $\mbhi$ or $\kappa$.
It is also useful to define the normalized scale factor $\tilde{a} \defeq a / \aRH$, where $a(t)$ is the scale factor with the normalization $a(t_0) = 1$ today and where $\aRH \defeq a(\taubh)$ is the scale factor at reheating (the instant of black-hole evaporation), which is calculated in Section~\ref{sec:reheating}.
Then $\tilde{a}(\tildet)$ is a universal function for black-hole domination, obtained by solving
\begin{equation}
    \dd{\tilde{a}}{\tildet} = \tilde{H}(\tilde{t}) \tilde{a}(\tilde{t}).
\end{equation}

Figure~\ref{fig:density-evol-universal} shows the evolution of the energy densities and the scale factor in black-hole domination.
For most of the lifetime of the black holes, the total energy density $\rhotot$ and scale factor $a$ remain close to those of a matter-dominated universe. However, near the end of the lifetime, these quantities deviate from the matter-dominated universe as Hawking radiation converts black-hole rest mass into relativistic energy, slowing the expansion of the universe.


\bibliographystyle{JHEP}
\bibliography{pbh-evaporation.bib}

\end{document}